\documentclass[sn-mathphys,Numbered]{sn-jnl}

\usepackage{graphicx}%
\usepackage{multirow}%
\usepackage{amsmath,amssymb,amsfonts}%
\usepackage{amsthm}%
\usepackage{mathrsfs}%
\usepackage[title]{appendix}%
\usepackage{xcolor}%
\usepackage{textcomp}%
\usepackage{manyfoot}%
\usepackage{booktabs}%
\usepackage{algorithm}%
\usepackage{algorithmicx}%
\usepackage{algpseudocode}%
\usepackage{listings}%

\theoremstyle{thmstyleone}%
\theoremstyle{thmstyletwo}%

\theoremstyle{thmstylethree}%
\newgeometry{left=2cm,bottom=1.5cm,top=1.2cm} 

\begin{document}

\title[Article Title]{Inducing mechanical self-healing in glassy polymer melts}


\author[1]{\fnm{Jos\'e} \sur{Ruiz-Franco}}

\author*[1]{\fnm{Andrea} \sur{Giuntoli}}\email{a.giuntoli@rug.nl}


\affil[1]{\orgdiv{Zernike Institute for Advanced Materials}, \orgname{University of Groningen}, \orgaddress{\postcode{9747 AG}, \state{Groningen}, \country{The Netherlands}}}



\abstract{Glassy polymer melts such as the plastics used in pipes, structural materials, and medical devices are ubiquitous in daily life. They accumulate damage over time due to their use, which limits their functionalities and demands periodic replacement. The resulting economic and social burden could be mitigated by the design of self-healing mechanisms that expand the lifespan of materials. However, the characteristic low molecular mobility in glassy polymer melts intrinsically limits the design of self-healing behavior. We demonstrate through numerical simulations that controlled oscillatory deformations enhance the local molecular mobility of glassy polymers without compromising their structural or mechanical stability. We apply this principle to increase the molecular mobility around the surface of a crack, inducing fracture repair and recovering the mechanical properties of the pristine material. Our findings establish a general physical mechanism of self-healing in glasses that may inspire the design and processing of new glassy materials.}





\maketitle

Self-healing refers to the ability of a material to restore functionality after undergoing physical damage. This characteristic is prominent in living materials such as plants~\cite{paris2007nitric}, mammalian cells~\cite{han2007dysferlin}, and human skin~\cite{chortos2016pursuing}. In contrast, synthetic materials lack self-healing mechanisms and the accumulated damage over time leads to macroscopic fractures and mechanical failure. In recent decades, the search for self-healing mechanisms in synthetic materials has become increasingly important for a sustainable environment~\cite{bekas2016self,liu2022healable}, promising to lower waste production and energy costs by diminishing the need for manufacturing replacements. Chemical modifications introducing dynamic covalent~\cite{terryn2017self,canadell2011self} and physical bonds~\cite{chen2012multiphase,urban2018key}, that break and reform reversibly can add self-healing abilities to synthetic materials. However, these mechanisms strongly rely on a sufficiently high molecular mobility within the material which facilitates mass transport, favoring the reconnection of crack surfaces and subsequent healing~\cite{wang2020self,hager2010self}.

Molecular mobility typically shows a dramatic decrease once a material is cooled below its glass-transition temperature $T_{g}$. Below $T_{g}$, materials exhibit rigidity because molecules are kinetically arrested. Above $T_{g}$, materials become flexible and viscous. Although soft materials such as elastomers and hydrogels maintain a solid shape due to covalent bonds, molecular segments have high local mobility at room temperature, i.e. $T_{g}<T_{r}$, which enables the use of intrinsic mechanisms to induce self-healing in soft materials~\cite{li2016highly,wu2017tough,jeon2016extremely,bilodeau2017self,fan2020advances}. In contrast, glassy polymer melts evidence an extremely low local mobility, since $T_{g} \gg T_{r}$. The kinetically arrested nature of glassy systems gives rise to mechanical stability and a high elastic modulus, but disables potential intrinsic mechanisms that induce self-healing in soft materials. To overcome this low mobility, alternative approaches rely on the synthesis of polymers with complex architectures or functionalization~\cite{yanagisawa2018mechanically,wang2020room,chen2012multiphase} that are not suited for large-scale production since the synthesis becomes complex, expensive and may give rise to mechanical properties different from those required~\cite{boils2004molecular,lebel2006dark,sheiko2019architectural,asadi2023tuning}. Therefore, these strategies apply only to very specific materials and create a critical bottleneck for the development of general self-healing glassy polymer melts such as everyday plastics.

Here we demonstrate by numerical simulations a general alternative approach that involves applying controlled oscillatory deformations on damaged glassy polymer melts to heal micro-cracks. Traditionally, deformations are associated with damage and fracture propagation. Large deformations in glassy materials trigger mechanical rejuvenation, restoring the system to an unaged state, similar to the thermodynamic effect of temperature increase. However, experimental studies on glassy polymers undergoing uniaxial deformation have shown enhanced molecular mobility before altering the underlying physical aging~\cite{lee2010mechanical,bennin2019enhanced}. Metallic polycrystalline alloys can undergo healing under well-controlled tensile high-cycle fatigue~\cite{barr2023autonomous} by merging the crack propagation front at the surface of grain boundaries. The orientation of the grain boundary planes arrests the crack propagation, promoting the healing via cold welding~\cite{wagle2015cold,dai2016side}. Likewise, numerical simulations of nanocrystalline metals reported that tensile and oscillatory shear deformations~\cite{xu2013healing} induce stress-driven grain boundary migration. We expect that similar mechanisms would also be observed in damaged amorphous glasses, which we prove numerically for model glassy polymer melts.

\section{Results}

Our framework involves glassy polymer melts with short chains to avoid entanglement effects. Beads belonging to the same chain interact via harmonic covalent bonds. Non-consecutive beads and beads from different chains interact attractively, corresponding to non-covalent interactions~\cite{giuntoli2020predictive}. The bead size is $d$, setting our length units (see the Methods section for a detailed model description). We explore a range of temperatures $T$ from a fluid $\left(T > T_{g}\right)$ to a glassy $\left(T < T_{g}\right)$ state. We quantify the molecular mobility from the $\alpha$-relaxation time $\tau_\alpha$ extracted from the decay of intermediate self-scattering function $F_{s}\left(q^{*},t\right)$, with $q^{*}$ corresponding to the length scale of the glassy cage, see Fig. S1.

\begin{figure}[!t]
\includegraphics[width=\linewidth]{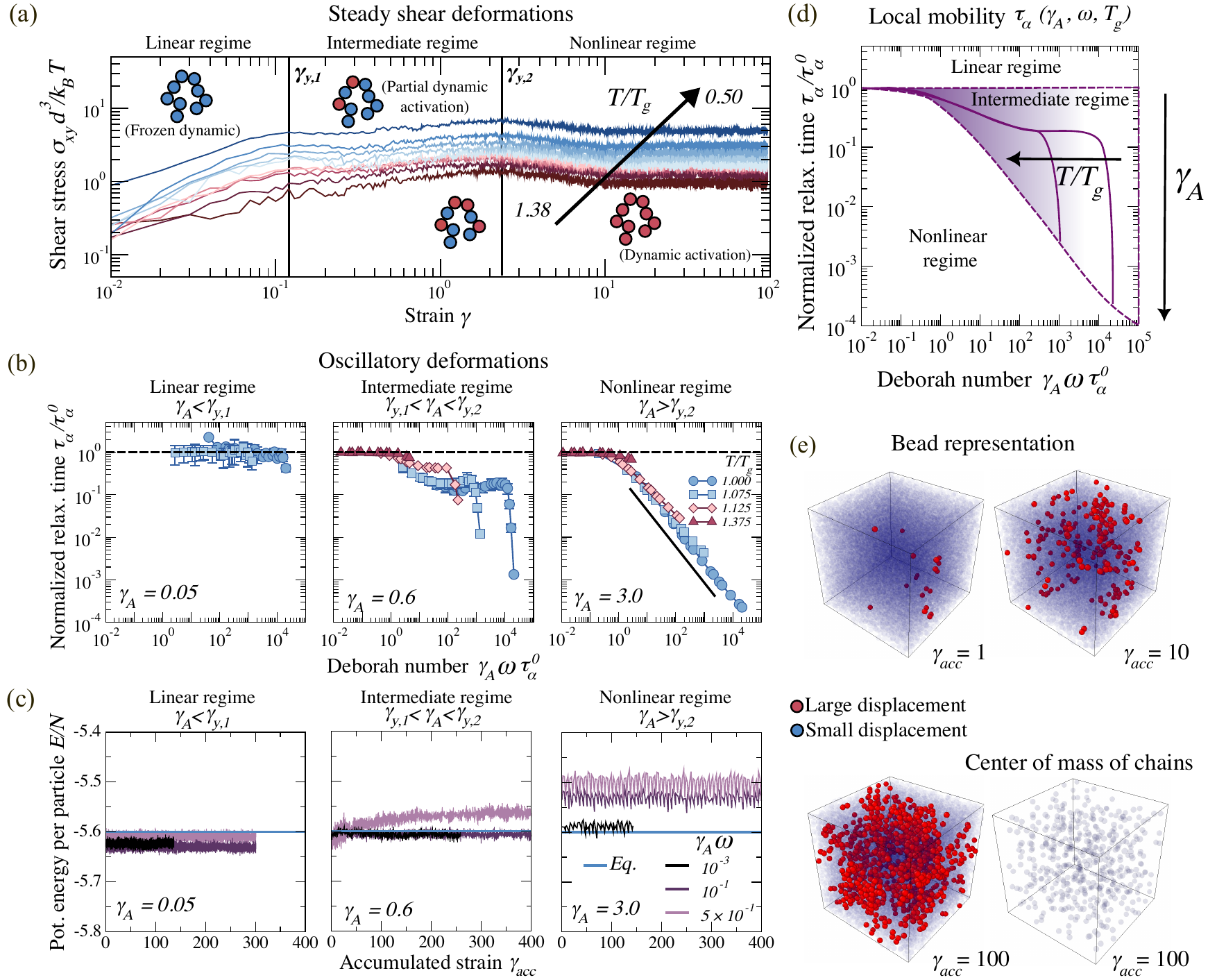}
\caption{\label{M1} \textbf{Quantifying bulk mobility}. (a) Shear stress tensor $\sigma_{xy}$ as a function of the strain $\gamma$ for a shear rate $\dot{\gamma}=0.01$ applied in the $xy-$plane. The arrow points to the decrease in temperature $T$. The first yielding point corresponding to physical bond breaking is referred to as $\gamma_{y,1}$, whereas $\gamma_{y,2}\simeq2.6$ indicates the yielding point corresponding to the transition from elastic solid to viscous-fluid behavior. The existence of these two yielding points delineates three regimes where all beads are frozen (linear regime, blue beads), partially relaxing (intermediate regime), or fully relaxing (nonlinear regime, red beads). (b) Normalized relaxation time $\tau_{\alpha}/\tau_{\alpha}^{0}$ as a function of the Deborah number $\gamma_{A}\omega\tau_{\alpha}^{0}$, for systems at different $T$ subjected to oscillatory deformations. (c) Energy per particle $E/N$ as a function of accumulated strain $\gamma_{acc}$ and $\gamma_{A}\omega$ for a system at $T/T_{g}=1.00$. (d) General representation of the local mobility under oscillatory shear deformation, $\tau_{\alpha}\left(\gamma_{A}, \omega, T \right)$. Black arrows emphasize how the structural relaxation time changes with increasing $\gamma_{A}$ and $T$ of the system. (e) Snapshots at different accumulated strain $\gamma_{acc}$ for a polymer melt system at $T/T_{g}=1.00$, subjected to an oscillatory deformation in the intermediate regime with $\gamma_{A}=0.4$ and $\gamma_{A}\omega = 0.01$. We also represent the displacement of the center of mass for the polymer chains. In this case, chains whose center of mass displaces equal to or more than the radius of gyration $R_{g}$ are not found.}
\end{figure}

Before showing that oscillatory deformations can heal fractures in glassy melts, we characterize the bulk $\alpha$-relaxation dynamics under shear. The oscillatory deformation depends on the strain amplitude $\gamma_{A}$ and frequency $\omega$, as $\gamma\left(t\right)=\gamma_{A}sin\left(\omega t\right)$. The combination of possible values within the set $\{\gamma_{A},\omega\}$ is vast, even restraining our simulations to a regime where the system temperature remains stable in the canonical $NVT$ ensemble. To restrict and rationalize our design space, we first study the rheological behavior of glassy polymer melts under steady shear deformations. The stress-strain evolution picture is well established in amorphous solid materials~\cite{ghosh2023microscopic}: there is an initial linear response corresponding to solid elastic behavior at small deformations dominated by sterical cages, while plastic deformations at large strains eventually trigger a viscous fluid behavior due to cage breaking. This transition is marked by a stress overshoot, placed at the yielding point $\gamma_{y,2}$. This picture is true for particle suspensions interacting sterically. However, in the present system, we consider linear chains interacting through attractive non-covalent bonds. As illustrated in Fig.~\ref{M1}(a), a first yielding point $\gamma_{y,1}$ emerges at small strain values as transient bonds break, allowing monomers to escape from interparticle attraction and partially dissipate energy~\cite{pham2008yielding,koumakis2011two,moghimi2020mechanisms}. In addition, Fig. S2 shows that $\gamma_{y,1}$ is shear rate dependent. The existence of two yielding points allows us to identify \textbf{three different regimes} of $\gamma_{A}$, with $\omega$ remaining a free parameter. Focusing on systems at $T/T_{g} \geq 1.0$, for which the polymer melt reaches equilibrium, we observe from Fig.~\ref{M1}(b) that $\gamma_{A}$ defined in each regime induces different molecular mobility. Both axes are rescaled by the relaxation time at equilibrium $\tau_{\alpha}^{0}$, so that relaxation times are reported in comparison to their equilibrium value, and deformation rates are reported as dimensionless Deborah numbers. This number defines the ratio of the material's characteristic relaxation time, in our case $\tau_{\alpha}^{0}$, to the characteristic flow time, that for oscillatory deformation is defined as $\gamma_{A}\omega$. The molecular mobility is extracted from $F_{s}\left(q^{*},t\right)$ under oscillatory deformation, see Fig. S3.

\underline{\textbf{Linear regime}} ($\gamma_{A} < \gamma_{y,1}$): oscillatory deformations have no effect on the $\alpha$-relaxation, as indicated by the flat behavior of $\tau_{\alpha}/\tau_{\alpha}^{0} \simeq 1$. However, the potential energy per particle $E/N$ as a function of the accumulated strain $\gamma_{acc}=4N_{cyc}\gamma_{A}$, shown in Fig.~\ref{M1}(c) and where $N_{cyc}$ represents the deformation cycles, is slightly smaller than the corresponding energy for the system immediately before applying oscillatory deformations. Previous numerical simulations of oscillatory athermal quasistatic shear (AQS) deformations of dense amorphous samples~\cite{regev2013onset,leishangthem2017yielding,schinasi2020annealing} established the limit at small strain deformations, i.e., $\gamma_{A}\ll \gamma_{y,2}$, as a way to obtain better-annealed glasses. In particular, AQS assumes that the relaxation after deformation occurs on shorter timescales than $\omega^{-1}$. Thus, deformations in the AQS approach depend on the deformation length scale encoded in the amplitude deformation. In our case, the deformation frequency $\omega$ introduces a time scale dependence that plays a key role in annealing glasses by accessing deeper energy states at small amplitude deformations~\cite{yeh2020glass,bhaumik2022yielding}. Although the study of this regime is beyond the scope of this investigation, we could anticipate that in the limit of poorly annealed glasses~\cite{yeh2020glass,bhaumik2022yielding}, and with decreasing $\omega$ (indicating that the deformation time scale is so large that it allows the system to relax the accumulated stress) our findings would align with the observations discussed in AQS simulations~\cite{parmar2019strain}.

\underline{\textbf{Non-linear regime}} ($\gamma_{A} > \gamma_{y,2}$): oscillatory deformations promote melting similarly to steady shear flow~\cite{giuntoli2020predictive}. In this regime, each oscillatory cycle induces a strain larger than the characteristic deformation that triggers the viscous-fluid transition. As a result, $\tau_{\alpha}$ decreases below $\tau_{\alpha}^{0}$ as $\gamma_{A}\omega$ increases through $\omega$. Actually, we find that that the slope of $\tau_{\alpha}/\tau_{\alpha}^{0}$ tends to an asymptotic behavior as $x^{-0.8}$ by decreasing $T$ and increasing the deformation. 
The exponent is related to the shear thinning behavior, which is characterized by an apparent viscosity that decreases with increasing shear rate~\cite{chen2011theory,giuntoli2020predictive}. Additionally, $E/N$ monotonically increases with $\gamma_{A}\omega$, shown in Fig.~\ref{M1}(c), whereas the characteristic plateau observed in $F_{s}\left(q^{*},t\right)$ shortened, see Fig. S3.

\underline{\textbf{Intermediate regime}} ($\gamma_{y,1} < \gamma_{A} < \gamma_{y,2}$):
intermediate strain amplitudes can speed up the molecular mobility without changing the underlying structure of the system. When the characteristic deformation time exceeds the structural relaxation time of the system, identified by Deborah numbers larger than $1$, the dynamics accelerates until it reaches a plateau. Maintaining a constant strain amplitude, we detect that the plateau length increases as we approach $T_{g}$. On the other hand, $\tau_{\alpha}$ falls to a plateau of lower height as $\gamma_{A}$ approaches $\gamma_{y,2}$, see Fig. S4. As $\gamma_{A}\omega$ increases by increasing $\omega$, the system approaches a critical value after which $\tau_{\alpha}$ resembles that of a polymer fluid. This striking behavior is in agreement with the enhanced molecular mobility observed for polymer glasses under tensile deformations prior to flow~\cite{lee2010mechanical,bennin2019enhanced}. Similarly, regions of high local mobility have been detected for amorphous materials under oscillatory shear deformations~\cite{parmar2019strain,yeh2020glass,bhaumik2021role,bhaumik2022avalanches}. These regions of high local mobility forming shear bands were identified at $\gamma_{A}>\gamma_{y,2}$, and we investigate if a similar mechanism is observed in our systems.

\begin{figure}[!t]
\includegraphics[width=\linewidth]{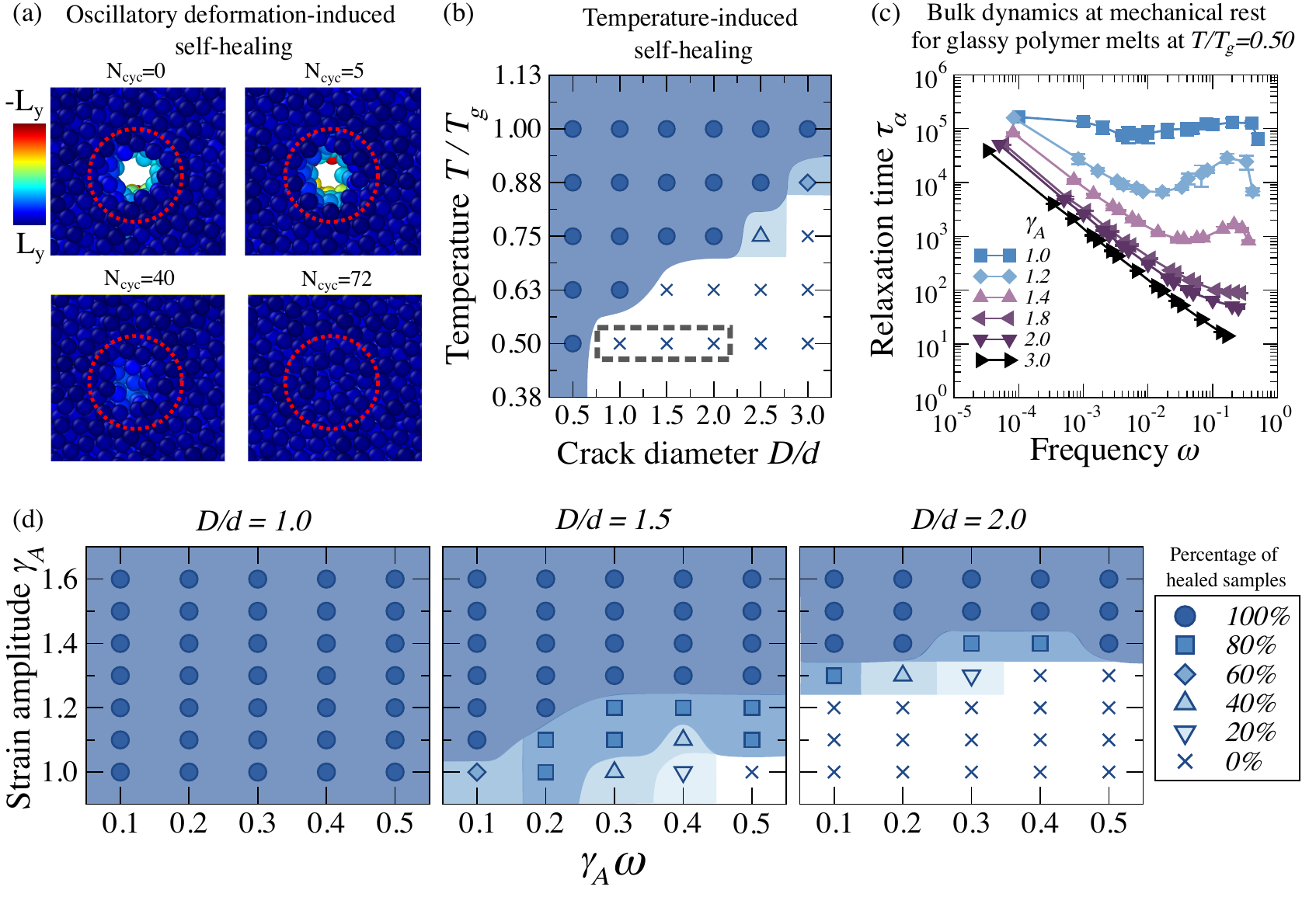}
\caption{\label{M2} \textbf{Healing crack in oscillating glasses}. (a) A cylindrical crack of diameter $D/d=1.0$ in a glassy polymer melt is repaired after several cycles of oscillatory deformations at $\gamma_{A}=1.0$ and $\gamma_{A}\omega=0.1$. Beads are colored according to their depth within the material. Dotted circles highlight the initial position of the cylindrical crack. (b) $\gamma_{A}-\gamma_{A}\omega$ phase diagram for healed glasses at different $D$ values of cylindrical cracks and varying strain amplitude and shear rate. (c) Structural relaxation time $\tau_{\alpha}$ as a function of amplitude $\gamma_{A}$ and frequency $\omega$ deformation. (d) $T-D$ phase diagram indicating the percentage of closed samples by increasing the overall system temperature. The dashed box highlights the range of cylindrical crack diameters we explore under the action of an oscillatory deformation.}
\end{figure}

One way to detect shear bands is by monitoring the particle energy $E/N$ as a function of $\gamma_{acc}$. As discussed in Ref.~\cite{parmar2019strain}, $E/N$ would exhibit a sharp upward shift because of the spontaneous onset of shear banding. In the intermediate and nonlinear regime, depicted in Fig.~\ref{M1}(c) and Fig. S5, we see that $E/N$ evolves smoothly without showing abrupt changes. Alternatively, we examine the mean-squared displacement $\left\langle r^{2} \right\rangle$ for single beads and for the center of mass of each chain as a function of $\gamma_{acc}$ (see Methods for more details). In Fig.~\ref{M1}(e) we identify beads that undergo a displacement $\left\langle r^{2} \right\rangle \geq d^{2}$, and chains whose center of mass undergo a displacement equal to or larger than the radius of gyration $R_{g}$, for a glassy system subjected to $\gamma_{A}=0.4$ and $\gamma_{A}\omega=10^{-1}$. Contrary to what has been observed in Refs.~\cite{parmar2019strain,yeh2020glass,ozawa2018random}, particles with high displacement do not develop a shear band. Furthermore, Fig. S6 shows that higher $\gamma_{A}$ promotes the formation of a branched percolating cluster of particles with a high displacement, resembling the scenario observed under steady shear~\cite{shrivastav2016yielding}. Therefore, the enhanced molecular mobility we observe does not involve shear banding. Instead, the existence of these plateaus suggests that in the intermediate amplitudes regime the oscillatory deformation breaks the microscopic cages responsible for the material vitrification. This mechanism, inducing local yielding, allows the system to remain globally arrested, but to relax locally. Fig.~\ref{M1}(d) schematizes molecular dynamics behavior under oscillatory deformation, expressed as $\tau_{\alpha} \left(\gamma_{A}, \omega, T \right)$.

The accelerated dynamics observed in Fig.~\ref{M1} can be leveraged to heal a crack in a glassy system via oscillatory deformations, as shown in Fig~\ref{M2}. As explained in the \textit{Methods} section, glassy polymer melt systems at $T/T_{g}=0.5$ are prepared with a cylindrical crack of diameter $D$ along the $y-$axis, with walls preventing particles from entering the crack region. Once the glassy state is reached, we remove the cylinder and allow the system to evolve freely. The first snapshot in Fig.~\ref{M2}(a) shows a glassy polymer melt system with a cylindrical crack of diameter $D/d=1.0$. The evolution of the crack surface area is monitored over time, shown in Fig. S8 and Supplementary Video 1 and 2, respectively. For cylindrical cracks of diameter half the particle size, i.e. 
$D/d=0.5$, the attractive interaction between different chains promotes inward flow. However, crack closure for $D/d \geq 1.0$ requires higher temperatures, see Fig.~\ref{M2}(b).

\begin{figure}[!t]
\includegraphics[width=\linewidth]{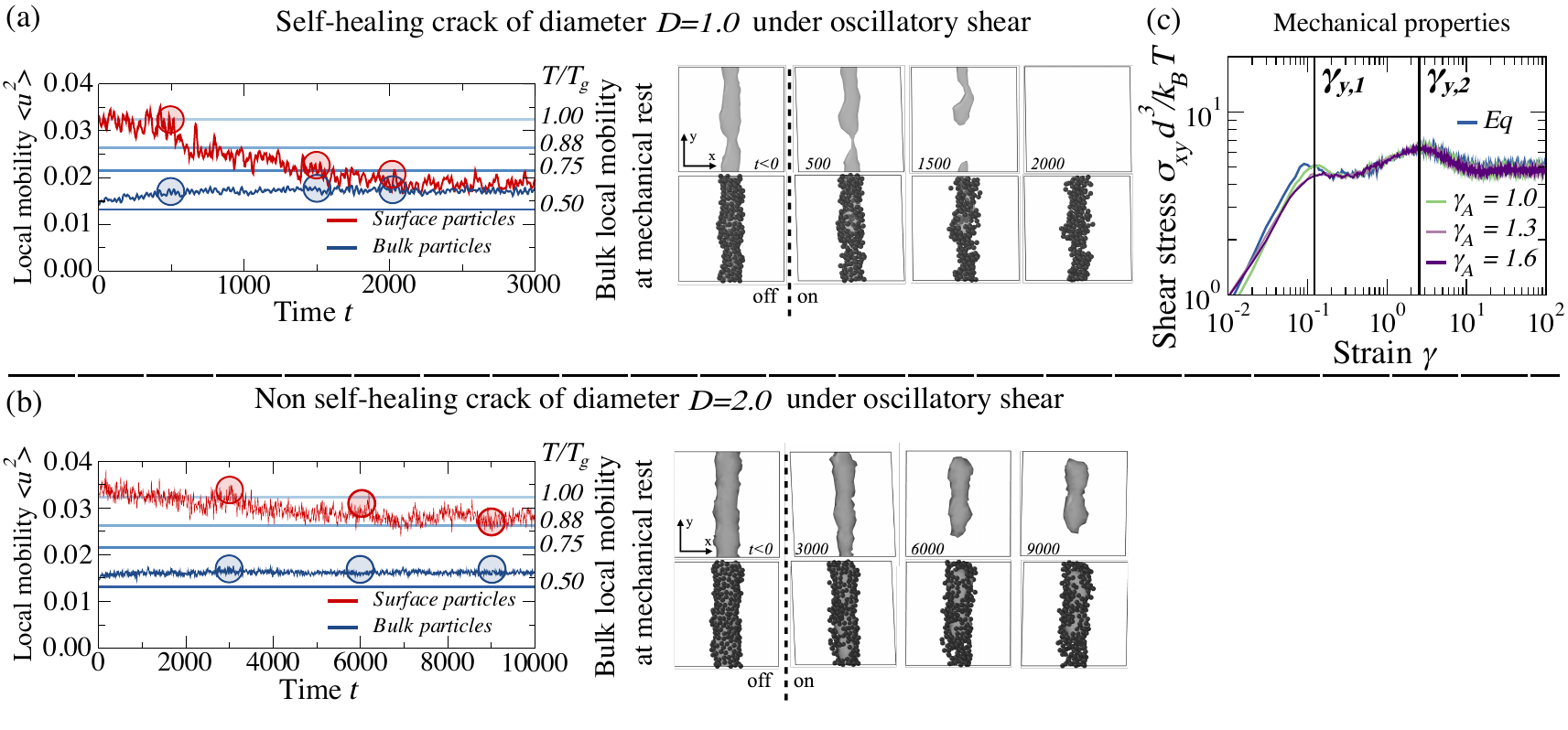}
\caption{\label{M3} \textbf{Local mobility and restoration of mechanical properties}. Local mobility $\left\langle u^{2}\right\rangle$ for particles on the crack surface and bulk as a function of time $t$ for glassy polymer melts subjected to oscillatory deformations of $\gamma_{A}=1.0$ and $\gamma_{A}\omega=0.1$, with a cylindrical crack of diameter (a) $D/d=1.0$ and (b) $D/d=2.0$. Horizontal lines represent the bulk local mobility as a function of $T$ at mechanical rest. Circles highlight the times at which the snapshots were taken, represented on the right. The crack region is represented in light grey, whereas dark grey beads correspond to beads on the crack surface. (c) The healed system at $\omega=0.1$ and different $\gamma_{A}$ values is subjected to steady shear flow at shear rate $\dot{\gamma}=0.01$. The resulting stress tensor is compared with the stress curve of the glass at rest, showing that mechanical properties are recovered.}
\end{figure}

Keeping $T/T_{g}=0.5$ (glass regime), we focus on the highlighted range shown in Fig.~\ref{M2}(b) and apply oscillatory deformations. Firstly, we repeat the previous study discussed in Fig.~\ref{M1}(b) by computing the evolution of $\tau_{\alpha}$ as a function of oscillatory deformations, see Fig.~\ref{M2}(c). While we cannot estimate $\tau_{\alpha}^{0}$ because the system is out-of-equilibrium (see Fig. S1), we observe qualitatively the same behavior as we approach $T_{g}$. Oscillations with $\gamma_{A}>\gamma_{y,2}$ lead to the melting of the glass material, whereas for $\gamma_{y,1}<\gamma_{A}<\gamma_{y,2}$, the dynamics can be locally accelerated by breaking microscopic cages. However, we also note that $\tau_{\alpha}$ develops a stark maximum by increasing $\omega$. This trend is already observed for $T\geq T_{g}$ (see intermediate regime in Fig. S4). The presence of this maximum following a sharp decrease in $\tau_{\alpha}$ could indicate non-trivial annealing behavior of the glasses, similar to that observed near the yielding point in Kob-Anderson binary mixtures~\cite{parmar2019strain}. However, in Fig. S7 we have attempted to establish a connection between $\tau_{\alpha}$ and the degree of molecular mobility by computing the distribution of per-particle Debye-Waller factor, but we do not see signs of any annealing. Instead, we observe that this peak marks the transition from a well-defined unimodal distribution to a distribution with a long tail. Returning to the mechanical self-healing, Fig.~\ref{M2}(a) displays the closure of a crack of $D/d=1.0$ after $N_{cyc}=72$ deformation cycles. Fig.~\ref{M2}(d) displays a set of $\gamma_{A}-\gamma_{A}\omega$ phase diagrams as a function of the initial $D$. These diagrams illustrate the percentage of samples in which oscillatory deformations in the intermediate regime successfully closed the crack. While for $D/d=1.0$ we consistently achieve crack closure, as $D$ increases the efficiency of oscillatory deformation decreases. However, we note that this decline begins at high values of $\omega$ for $D/d=1.5$, and becomes more pronounced for $D/d=2.0$. This trend is rationalized by the fact that the oscillation rate of the deformation imposed by $\omega$ overcomes the characteristic particle diffusion rate. This, along with the fact that the crack diameter is large enough to suppress particle interactions across the crack, favors particles following the same trajectory in each deformation cycle. Thus, an increase in $\gamma_{A}$ is required to bring particles in contact across the crack.

One may wonder if we genuinely induce self-healing by accelerating the local dynamics exclusively around the crack. This is a legitimate doubt because deformations are often associated with the breakdown of the material's structure inducing mechanical rejuvenation. In addition to showing that this amplitude regime precedes the material's yielding point $\gamma_{y,2}$, Fig.~\ref{M3} shows that indeed cracks can heal purely by accelerated local dynamics, while the bulk remains globally glassy and the final mechanical properties of the healed glass are comparable to those of the pristine equilibrium glass. We identify particles on the surface of the crack and in the bulk, i.e. far away from the crack, and compute their mobility during deformation through the Debye-Waller factor $\left\langle u^{2} \right\rangle$, which corresponds to the value of the mean square displacement at local minima of its logarithmic derivative~\cite{larini2008universal,zhu2022effect}, as shown in Fig. S1(e). Fig.~\ref{M3}(a) displays results corresponding to a glassy polymer melt with a crack of diameter $D/d=1.0$, subjected to $\{\gamma_{A},\omega\}=\{1.0,0.1\}$, equivalent to $\gamma_{A}\omega=0.1$. The crack surface and particles on the surface are represented in the snapshots in light grey and dark grey colors, respectively, while particles in the bulk are omitted to enhance the clarity of the snapshots. Furthermore, we depict the evolution of $\left\langle u^{2} \right\rangle$ as a function of the time $t$ alongside the corresponding mobility at different $T$ in the absence of deformation, indicated by a horizontal line. While $\left<u^{2}\right>$ slightly increases for particles in the bulk, we note that the dynamics for surface particles aligns with that observed for a bulk equilibrium system at $T/T_{g}=1.0$. As the crack closes, 
the surface dynamics converges to the bulk dynamics, at a value slightly higher than the equilibrium dynamics at $T/T_{g}=0.50$ because of the oscillatory deformations. It is important to note how the particles initially located on the surface remain localized around the region where the crack was. This indicates that local melting induces an inward flow of material, filling the damaged space. In the case where the crack does not close, as depicted in Fig.~\ref{M3}(b), the surface and bulk dynamics do not match.

Finally, we assess the mechanical properties of the material immediately after closing the crack. In Fig.~\ref{M3}(c), we compare the mechanical response under steady shear of the healing glass material with those obtained from a glassy polymer melt in which no crack is present. At low-strain deformations, the first yielding point $\gamma_{y,1}$, associated with physical bond breaking, is less apparent in the stress curves of the healed material. This issue is attributed to the slightly accelerated bulk dynamics, see Fig.~\ref{M3}(a) and Fig. S9. However, the position of $\gamma_{y,2}$ and the stress overshoot $\sigma_{xy}\left(\gamma_{y,2}\right)$, indicating the opposing resistance of the material to break and flow, perfectly match. Thus, our findings suggest that the application of small oscillatory deformations has the potential to induce complete self-healing in glassy polymer materials.

\section{Discussion}
In recent decades, oscillatory deformations have emerged as an avenue for exploring new frontiers in the fundamental physics of amorphous materials. This includes enhancing annealing~\cite{yeh2020glass,schinasi2020annealing} and inducing memory~\cite{fiocco2014encoding,adhikari2018memory} in amorphous glasses. Here, we used oscillatory deformations, to induce self-healing in glassy polymer materials by overcoming their intrinsically low local mobility. The enhanced mobility would enable the use of intrinsic healing mechanisms such as dynamic covalent bonds or supramolecular bonds that strongly depend on the local dynamics~\cite{urban2018key}. 

First, we have studied the local mobility in bulk model glasses without damage. We have identified a range of oscillatory deformations where the molecular mobility can be meticulously accelerated without globally modifying the underlying structure or mechanical properties of the system. Then we have demonstrated that in this range, oscillatory deformations can stimulate crack closure and damage repair by accelerating the dynamics of particles around the crack surface, while the bulk dynamics is only slightly perturbed. Finally, we have also shown that the stress curves of the healed samples closely match the pristine material.

Since glassy polymers are out-of-equilibrium materials, their rheological response depends on the aging history and preparation protocol. In the past, it has been studied that more stable glasses are more brittle, making the yielding point a spinodal instability characterized by a sharp discontinuous stress jump~\cite{ozawa2018random}. It would be relevant in the future to understand if local mobility controlled by oscillatory deformations depends on the glass stability. Likewise, crack geometry in fractured material can acquire erratic shapes and orientations. Exploring local mobility on erratic crack surfaces would also be another relevant point. Nevertheless, our results on glassy polymer melts, along with previous observations of healing in metal alloys~\cite{barr2023autonomous} and nanocrystalline metals~\cite{xu2013healing} under mechanical deformations, point to the fact that oscillatory deformations represent a general strategy to induce mechanical self-healing in materials with extremely low local mobility.

\section{Methods}\label{secMet}
\textbf{Modelling.} We perform Molecular Dynamics simulations of fully flexible linear chains of beads linked by harmonic springs. Nonbonded monomers belonging to the same or different chains interact with a truncated LJ potential defined as

\begin{equation}
U_{LJ}\left(r \right )=\left\{\begin{matrix}
4\epsilon\left[ \left(\frac{d}{r}\right)^{12} - \left(\frac{d}{r}\right)^{6}  \right] + U_{cut} & \mathrm{if} & r\leq2.5d \\ 
0 & \mathrm{if} & r>2.5d
\end{matrix}\right.\,,
\label{eq:LJ}
\end{equation}

\noindent where $d$ is the particle diameter, which sets the unit of length, $\epsilon$ controls the energy scale, and $U_{cut}$ ensures that $U_{LJ}\left(r=2.5d \right )=0$. Defining $m$ as the mass of the particles, the time units are defined as $t=\sqrt{md^{2}/\epsilon}$. On the other hand, chemical bonds between connected monomers are modeled by 

\begin{equation}
    U_{b}\left(r\right)=k_{b}\left(r-r_{0}\right)^{2}\,,
\label{eq:Bond}
\end{equation}

\noindent where $k_{b}=555.5\epsilon/d^{2}$ is the spring constant and $r_{0}=0.97d$ is the equilibrium bond length. Henceforth, all quantities are expressed in terms of reduced LJ units, i.e., $\epsilon=1$ and $d=1$, with unit monomer mass $m$ and Boltzmann constant. The reduced units can be mapped onto physical units relevant to generic nonequilibrium fluids, by taking molecular dynamics (MD) time, length, and
energy units as corresponding roughly to about $2$ ps, $0.5$ nm, and
$3.7$ kJ/mol, respectively. 

We consider systems of $N_{c}=500$ chains of $M=20$ monomers, avoiding thus the entanglement, and doing a total of $N=10^{4}$ monomers. All simulations were performed using LAMMPS simulation package \cite{LAMMPS}, considering a simulation time step $\delta t=0.002$. Results are averaged over 5 independent simulations.

\noindent\textbf{Sample preparation.} Initially, an equilibration process was performed by considering polymer chains enclosed in an orthogonal cubic box of size $L$ with periodic boundary conditions. First, $10^{7}$ simulation steps were performed in the NPT ensemble by employing a Nos\'e-Hoover thermostat and barostat with $T=1.0$ and $\left\langle P \right\rangle = 0$ to allow full correlation loss of the end-to-end vector of the polymer chains. Next, a quenching was made in the range of temperatures from $T=0.20$ to $T=0.55$, maintaining $\left\langle P \right\rangle = 0$ during $10^{7}$ simulation steps. In both equilibration procedures, the simulation box was allowed to fluctuate in an isotropic fashion in all three directions of space.

\noindent\textbf{Static and dynamics characterization.} Once the system was equilibrated to the required temperature, the volume was fixed by performing simulations in the \textit{NVT} ensemble by using a Nos\'e-Hoover thermostat at the required temperature, and static and dynamic properties were computed to characterize the polymer system as a function of $T$. In particular, the static structure factor was computed as $ S\left(q\right)= \left\langle \frac{1}{N} \sum_{ij} e^{ -i\mathbf{q}\cdot\left( \mathbf{r}_{i}-\mathbf{r}_j \right)}  \right\rangle\ $, where $\mathbf{q}$ is the wave vector, $\mathbf{r}_{i}$ indicates the position of the $i-$th particle, and $\left\langle \cdots \right\rangle$ denotes an ensemble average. Likewise, the dynamic of the system was quantified through the mean-squared displacement $\left\langle \Delta r^{2}\left(t\right) \right\rangle = \left\langle \frac{1}{N} \sum_{i=1}^{N} \left[ \mathbf{r}_{i}\left(t\right) - \mathbf{r}_{i}\left(0\right)  \right]^{2} \right\rangle$ and the self-intermediate scattering function $ F_{s}\left(q^{*},t\right) = \left\langle \frac{1}{N}  \sum_{i}^{N} e^{i\mathbf{q}^{*} \cdot \left[ \mathbf{r}_{i}\left(t+t'\right) - \mathbf{r}_{i}\left(t'\right) \right]}\right\rangle $, computed on the wavevector $q^{*}=2\pi/l$ corresponding to the main peak of the static structure factor. In those cases where the system recovered ergodicity, i.e. $F_{s}\left(q^{*},t\right)=0$ in the simulated time window, the relaxation time $\tau_{\alpha}$ was extracted by imposing $F_{s}\left(q^{*},\tau_{\alpha}\right)=1/e$.

\noindent\textbf{Shear deformation.} After equilibration, we perform two kinds of simulations: (i) steady shear deformations at a fixed shear rate $\dot{\gamma}$; (ii) oscillatory shear deformations by applying a sinusoidal deformation in the $xy$ plane defined as $\gamma\left(t\right)=\gamma_{A}sin\left(\omega t\right)$, where $\gamma_{A}$ is the strain amplitude and $\omega$ is the frequency. In both cases, deformations are performed at constant volume and temperature, and hence, the SLLOD equations of motion were used along with the thermostat, in combination with Lees-Edwards boundary conditions. During the steady shear deformation, we monitor the stress tensor component $\sigma_{\alpha\beta} = \left\langle  \frac{1}{A} \sum_{ij} f_{ij}^{\alpha} r_{ij}^{\beta} \right\rangle$, where $f_{ij}^{\alpha}$ is the $\alpha-$component of the force with respect to interaction defined in eq. (\ref{eq:LJ}) and eq. (\ref{eq:Bond}), $r_{ij}^{\beta}$ corresponds the $\beta-$component of the distance vector between the particles $i$ and $j$. In this case, the steady shear deformation was applied in all three directions of space and subsequently averaged. On the other hand, during the oscillatory deformation, we first apply an accumulated strain $\gamma = 4N_{cyc}\gamma_{A} = 100\%$, where $N_{cyc}$ corresponds to the number of deformation cycles. Then, we compute the self-intermediate scattering function on the wave vector $q^{*}$ corresponding to the main peak of the static structure factor before applying any deformation. Since oscillatory deformations were applied on the $xy$ plane, $F_{s}\left(q^{*},t\right)$ was computed excluding $q_{x}$ components, see Fig. S3. Furthermore, we computed the mean-squared squared displacement per particle, defined as $ \Delta r^{2}\left(t\right)_{i} = \left[ \mathbf{r}_{i}\left(t\right) - \mathbf{r}_{i}\left(0\right)  \right]^{2}$, excluding the $x-$component.

\noindent\textbf{Inducing self-healing.} To study self-healing, we create a crack on glassy polymer melts with $T/T_{g}=0.50$ and $N$ monomers. The equilibration previously described above is performed in the presence of a cylinder placed at the center of the simulation box of length $L_{y}$ and diameter $D$. Monomers interacted with the cylinder wall through the Weeks-Chandler Andersen potential:

\begin{equation}
U_{WCA}\left(r \right )=\left\{\begin{matrix}
4\epsilon\left[ \left(\frac{d}{r}\right)^{12} - \left(\frac{d}{r}\right)^{6}  \right] + \epsilon & \mathrm{if} & r\leq2^{1/6}d \\ 
0 & \mathrm{if} & r>2^{1/6}d
\end{matrix}\right.\,.
\end{equation}

\noindent Then, the cylinder is removed, and NVT simulations and oscillatory deformation simulations are independently performed during $10^{7}$ up to $3\times10^{7}$ simulation steps. In both cases, the volume of the crack is monitored by employing the surface area algorithm \cite{stukowski2009visualization}. During both simulations, we compute the mobility $\left\langle u^{2} \right\rangle$ of the monomers as a function of the distance from the center of the simulation box, as well as the function of time. Since the mobility is statistically very noisy, we apply a Savitzky-Golay filter on windows of 21 points and equations of 2 order.

\noindent\textbf{Mechanical properties after the self-healing.} Glassy polymer systems, on which the crack was closed, are subjected to steady shear flow at $\dot{\gamma} = 0.01$, and the evolution of $\sigma_{xy}$ as a function of $\gamma=\dot{\gamma}t$ is compared with the corresponding $\sigma_{xy}$ for the glass at rest.

\newpage

\bibliography{sn-bibliography}


\begin{thebibliography}{50}
\ifx \bisbn   \undefined \def \bisbn  #1{ISBN #1}\fi
\ifx \binits  \undefined \def \binits#1{#1}\fi
\ifx \bauthor  \undefined \def \bauthor#1{#1}\fi
\ifx \batitle  \undefined \def \batitle#1{#1}\fi
\ifx \bjtitle  \undefined \def \bjtitle#1{#1}\fi
\ifx \bvolume  \undefined \def \bvolume#1{\textbf{#1}}\fi
\ifx \byear  \undefined \def \byear#1{#1}\fi
\ifx \bissue  \undefined \def \bissue#1{#1}\fi
\ifx \bfpage  \undefined \def \bfpage#1{#1}\fi
\ifx \blpage  \undefined \def \blpage #1{#1}\fi
\ifx \burl  \undefined \def \burl#1{\textsf{#1}}\fi
\ifx \doiurl  \undefined \def \doiurl#1{\url{https://doi.org/#1}}\fi
\ifx \betal  \undefined \def \betal{\textit{et al.}}\fi
\ifx \binstitute  \undefined \def \binstitute#1{#1}\fi
\ifx \binstitutionaled  \undefined \def \binstitutionaled#1{#1}\fi
\ifx \bctitle  \undefined \def \bctitle#1{#1}\fi
\ifx \beditor  \undefined \def \beditor#1{#1}\fi
\ifx \bpublisher  \undefined \def \bpublisher#1{#1}\fi
\ifx \bbtitle  \undefined \def \bbtitle#1{#1}\fi
\ifx \bedition  \undefined \def \bedition#1{#1}\fi
\ifx \bseriesno  \undefined \def \bseriesno#1{#1}\fi
\ifx \blocation  \undefined \def \blocation#1{#1}\fi
\ifx \bsertitle  \undefined \def \bsertitle#1{#1}\fi
\ifx \bsnm \undefined \def \bsnm#1{#1}\fi
\ifx \bsuffix \undefined \def \bsuffix#1{#1}\fi
\ifx \bparticle \undefined \def \bparticle#1{#1}\fi
\ifx \barticle \undefined \def \barticle#1{#1}\fi
\bibcommenthead
\ifx \bconfdate \undefined \def \bconfdate #1{#1}\fi
\ifx \botherref \undefined \def \botherref #1{#1}\fi
\ifx \url \undefined \def \url#1{\textsf{#1}}\fi
\ifx \bchapter \undefined \def \bchapter#1{#1}\fi
\ifx \bbook \undefined \def \bbook#1{#1}\fi
\ifx \bcomment \undefined \def \bcomment#1{#1}\fi
\ifx \oauthor \undefined \def \oauthor#1{#1}\fi
\ifx \citeauthoryear \undefined \def \citeauthoryear#1{#1}\fi
\ifx \endbibitem  \undefined \def \endbibitem {}\fi
\ifx \bconflocation  \undefined \def \bconflocation#1{#1}\fi
\ifx \arxivurl  \undefined \def \arxivurl#1{\textsf{#1}}\fi
\csname PreBibitemsHook\endcsname

\bibitem[\protect\citeauthoryear{Par{\'\i}s et~al.}{2007}]{paris2007nitric}
\begin{barticle}
\bauthor{\bsnm{Par{\'\i}s}, \binits{R.}},
\bauthor{\bsnm{Lamattina}, \binits{L.}},
\bauthor{\bsnm{Casalongu{\'e}}, \binits{C.A.}}:
\batitle{Nitric oxide promotes the wound-healing response of potato leaflets}.
\bjtitle{Plant Physiology and Biochemistry}
\bvolume{45}(\bissue{1}),
\bfpage{80}--\blpage{86}
(\byear{2007})
\end{barticle}
\endbibitem

\bibitem[\protect\citeauthoryear{Han and Campbell}{2007}]{han2007dysferlin}
\begin{barticle}
\bauthor{\bsnm{Han}, \binits{R.}},
\bauthor{\bsnm{Campbell}, \binits{K.P.}}:
\batitle{Dysferlin and muscle membrane repair}.
\bjtitle{Current opinion in cell biology}
\bvolume{19}(\bissue{4}),
\bfpage{409}--\blpage{416}
(\byear{2007})
\end{barticle}
\endbibitem

\bibitem[\protect\citeauthoryear{Chortos et~al.}{2016}]{chortos2016pursuing}
\begin{barticle}
\bauthor{\bsnm{Chortos}, \binits{A.}},
\bauthor{\bsnm{Liu}, \binits{J.}},
\bauthor{\bsnm{Bao}, \binits{Z.}}:
\batitle{Pursuing prosthetic electronic skin}.
\bjtitle{Nature materials}
\bvolume{15}(\bissue{9}),
\bfpage{937}--\blpage{950}
(\byear{2016})
\end{barticle}
\endbibitem

\bibitem[\protect\citeauthoryear{Bekas et~al.}{2016}]{bekas2016self}
\begin{barticle}
\bauthor{\bsnm{Bekas}, \binits{D.}},
\bauthor{\bsnm{Tsirka}, \binits{K.}},
\bauthor{\bsnm{Baltzis}, \binits{D.}},
\bauthor{\bsnm{Paipetis}, \binits{A.S.}}:
\batitle{Self-healing materials: A review of advances in materials, evaluation,
  characterization and monitoring techniques}.
\bjtitle{Composites Part B: Engineering}
\bvolume{87},
\bfpage{92}--\blpage{119}
(\byear{2016})
\end{barticle}
\endbibitem

\bibitem[\protect\citeauthoryear{Liu et~al.}{2022}]{liu2022healable}
\begin{barticle}
\bauthor{\bsnm{Liu}, \binits{X.}},
\bauthor{\bsnm{Li}, \binits{Y.}},
\bauthor{\bsnm{Fang}, \binits{X.}},
\bauthor{\bsnm{Zhang}, \binits{Z.}},
\bauthor{\bsnm{Li}, \binits{S.}},
\bauthor{\bsnm{Sun}, \binits{J.}}:
\batitle{Healable and recyclable polymeric materials with high mechanical
  robustness}.
\bjtitle{ACS Materials Letters}
\bvolume{4}(\bissue{4}),
\bfpage{554}--\blpage{571}
(\byear{2022})
\end{barticle}
\endbibitem

\bibitem[\protect\citeauthoryear{Terryn et~al.}{2017}]{terryn2017self}
\begin{barticle}
\bauthor{\bsnm{Terryn}, \binits{S.}},
\bauthor{\bsnm{Brancart}, \binits{J.}},
\bauthor{\bsnm{Lefeber}, \binits{D.}},
\bauthor{\bsnm{Van~Assche}, \binits{G.}},
\bauthor{\bsnm{Vanderborght}, \binits{B.}}:
\batitle{Self-healing soft pneumatic robots}.
\bjtitle{Science Robotics}
\bvolume{2}(\bissue{9}),
\bfpage{4268}
(\byear{2017})
\end{barticle}
\endbibitem

\bibitem[\protect\citeauthoryear{Canadell et~al.}{2011}]{canadell2011self}
\begin{barticle}
\bauthor{\bsnm{Canadell}, \binits{J.}},
\bauthor{\bsnm{Goossens}, \binits{H.}},
\bauthor{\bsnm{Klumperman}, \binits{B.}}:
\batitle{Self-healing materials based on disulfide links}.
\bjtitle{Macromolecules}
\bvolume{44}(\bissue{8}),
\bfpage{2536}--\blpage{2541}
(\byear{2011})
\end{barticle}
\endbibitem

\bibitem[\protect\citeauthoryear{Chen et~al.}{2012}]{chen2012multiphase}
\begin{barticle}
\bauthor{\bsnm{Chen}, \binits{Y.}},
\bauthor{\bsnm{Kushner}, \binits{A.M.}},
\bauthor{\bsnm{Williams}, \binits{G.A.}},
\bauthor{\bsnm{Guan}, \binits{Z.}}:
\batitle{Multiphase design of autonomic self-healing thermoplastic elastomers}.
\bjtitle{Nature chemistry}
\bvolume{4}(\bissue{6}),
\bfpage{467}--\blpage{472}
(\byear{2012})
\end{barticle}
\endbibitem

\bibitem[\protect\citeauthoryear{Urban et~al.}{2018}]{urban2018key}
\begin{barticle}
\bauthor{\bsnm{Urban}, \binits{M.W.}},
\bauthor{\bsnm{Davydovich}, \binits{D.}},
\bauthor{\bsnm{Yang}, \binits{Y.}},
\bauthor{\bsnm{Demir}, \binits{T.}},
\bauthor{\bsnm{Zhang}, \binits{Y.}},
\bauthor{\bsnm{Casabianca}, \binits{L.}}:
\batitle{Key-and-lock commodity self-healing copolymers}.
\bjtitle{Science}
\bvolume{362}(\bissue{6411}),
\bfpage{220}--\blpage{225}
(\byear{2018})
\end{barticle}
\endbibitem

\bibitem[\protect\citeauthoryear{Wang and Urban}{2020}]{wang2020self}
\begin{barticle}
\bauthor{\bsnm{Wang}, \binits{S.}},
\bauthor{\bsnm{Urban}, \binits{M.W.}}:
\batitle{Self-healing polymers}.
\bjtitle{Nature Reviews Materials}
\bvolume{5}(\bissue{8}),
\bfpage{562}--\blpage{583}
(\byear{2020})
\end{barticle}
\endbibitem

\bibitem[\protect\citeauthoryear{Hager et~al.}{2010}]{hager2010self}
\begin{barticle}
\bauthor{\bsnm{Hager}, \binits{M.D.}},
\bauthor{\bsnm{Greil}, \binits{P.}},
\bauthor{\bsnm{Leyens}, \binits{C.}},
\bauthor{\bsnm{Van Der~Zwaag}, \binits{S.}},
\bauthor{\bsnm{Schubert}, \binits{U.S.}}:
\batitle{Self-healing materials}.
\bjtitle{Advanced Materials}
\bvolume{22}(\bissue{47}),
\bfpage{5424}--\blpage{5430}
(\byear{2010})
\end{barticle}
\endbibitem

\bibitem[\protect\citeauthoryear{Li et~al.}{2016}]{li2016highly}
\begin{barticle}
\bauthor{\bsnm{Li}, \binits{C.-H.}},
\bauthor{\bsnm{Wang}, \binits{C.}},
\bauthor{\bsnm{Keplinger}, \binits{C.}},
\bauthor{\bsnm{Zuo}, \binits{J.-L.}},
\bauthor{\bsnm{Jin}, \binits{L.}},
\bauthor{\bsnm{Sun}, \binits{Y.}},
\bauthor{\bsnm{Zheng}, \binits{P.}},
\bauthor{\bsnm{Cao}, \binits{Y.}},
\bauthor{\bsnm{Lissel}, \binits{F.}},
\bauthor{\bsnm{Linder}, \binits{C.}}, \betal:
\batitle{A highly stretchable autonomous self-healing elastomer}.
\bjtitle{Nature chemistry}
\bvolume{8}(\bissue{6}),
\bfpage{618}--\blpage{624}
(\byear{2016})
\end{barticle}
\endbibitem

\bibitem[\protect\citeauthoryear{Wu et~al.}{2017}]{wu2017tough}
\begin{barticle}
\bauthor{\bsnm{Wu}, \binits{J.}},
\bauthor{\bsnm{Cai}, \binits{L.-H.}},
\bauthor{\bsnm{Weitz}, \binits{D.A.}}:
\batitle{Tough self-healing elastomers by molecular enforced integration of
  covalent and reversible networks}.
\bjtitle{Advanced materials}
\bvolume{29}(\bissue{38}),
\bfpage{1702616}
(\byear{2017})
\end{barticle}
\endbibitem

\bibitem[\protect\citeauthoryear{Jeon et~al.}{2016}]{jeon2016extremely}
\begin{barticle}
\bauthor{\bsnm{Jeon}, \binits{I.}},
\bauthor{\bsnm{Cui}, \binits{J.}},
\bauthor{\bsnm{Illeperuma}, \binits{W.R.}},
\bauthor{\bsnm{Aizenberg}, \binits{J.}},
\bauthor{\bsnm{Vlassak}, \binits{J.J.}}:
\batitle{Extremely stretchable and fast self-healing hydrogels}.
\bjtitle{Advanced Materials}
\bvolume{28}(\bissue{23}),
\bfpage{4678}--\blpage{4683}
(\byear{2016})
\end{barticle}
\endbibitem

\bibitem[\protect\citeauthoryear{Bilodeau and Kramer}{2017}]{bilodeau2017self}
\begin{barticle}
\bauthor{\bsnm{Bilodeau}, \binits{R.A.}},
\bauthor{\bsnm{Kramer}, \binits{R.K.}}:
\batitle{Self-healing and damage resilience for soft robotics: A review}.
\bjtitle{Frontiers in Robotics and AI}
\bvolume{4},
\bfpage{48}
(\byear{2017})
\end{barticle}
\endbibitem

\bibitem[\protect\citeauthoryear{Fan et~al.}{2020}]{fan2020advances}
\begin{barticle}
\bauthor{\bsnm{Fan}, \binits{L.}},
\bauthor{\bsnm{Ge}, \binits{X.}},
\bauthor{\bsnm{Qian}, \binits{Y.}},
\bauthor{\bsnm{Wei}, \binits{M.}},
\bauthor{\bsnm{Zhang}, \binits{Z.}},
\bauthor{\bsnm{Yuan}, \binits{W.-E.}},
\bauthor{\bsnm{Ouyang}, \binits{Y.}}:
\batitle{Advances in synthesis and applications of self-healing hydrogels}.
\bjtitle{Frontiers in Bioengineering and Biotechnology}
\bvolume{8},
\bfpage{654}
(\byear{2020})
\end{barticle}
\endbibitem

\bibitem[\protect\citeauthoryear{Yanagisawa
  et~al.}{2018}]{yanagisawa2018mechanically}
\begin{barticle}
\bauthor{\bsnm{Yanagisawa}, \binits{Y.}},
\bauthor{\bsnm{Nan}, \binits{Y.}},
\bauthor{\bsnm{Okuro}, \binits{K.}},
\bauthor{\bsnm{Aida}, \binits{T.}}:
\batitle{Mechanically robust, readily repairable polymers via tailored
  noncovalent cross-linking}.
\bjtitle{Science}
\bvolume{359}(\bissue{6371}),
\bfpage{72}--\blpage{76}
(\byear{2018})
\end{barticle}
\endbibitem

\bibitem[\protect\citeauthoryear{Wang et~al.}{2020}]{wang2020room}
\begin{barticle}
\bauthor{\bsnm{Wang}, \binits{H.}},
\bauthor{\bsnm{Liu}, \binits{H.}},
\bauthor{\bsnm{Cao}, \binits{Z.}},
\bauthor{\bsnm{Li}, \binits{W.}},
\bauthor{\bsnm{Huang}, \binits{X.}},
\bauthor{\bsnm{Zhu}, \binits{Y.}},
\bauthor{\bsnm{Ling}, \binits{F.}},
\bauthor{\bsnm{Xu}, \binits{H.}},
\bauthor{\bsnm{Wu}, \binits{Q.}},
\bauthor{\bsnm{Peng}, \binits{Y.}}, \betal:
\batitle{Room-temperature autonomous self-healing glassy polymers with
  hyperbranched structure}.
\bjtitle{Proceedings of the National Academy of Sciences}
\bvolume{117}(\bissue{21}),
\bfpage{11299}--\blpage{11305}
(\byear{2020})
\end{barticle}
\endbibitem

\bibitem[\protect\citeauthoryear{Boils et~al.}{2004}]{boils2004molecular}
\begin{barticle}
\bauthor{\bsnm{Boils}, \binits{D.}},
\bauthor{\bsnm{Perron}, \binits{M.-{\`E}.}},
\bauthor{\bsnm{Monchamp}, \binits{F.}},
\bauthor{\bsnm{Duval}, \binits{H.}},
\bauthor{\bsnm{Maris}, \binits{T.}},
\bauthor{\bsnm{Wuest}, \binits{J.D.}}:
\batitle{Molecular tectonics. disruption of self-association in melts derived
  from hydrogen-bonded solids}.
\bjtitle{Macromolecules}
\bvolume{37}(\bissue{19}),
\bfpage{7351}--\blpage{7357}
(\byear{2004})
\end{barticle}
\endbibitem

\bibitem[\protect\citeauthoryear{Lebel et~al.}{2006}]{lebel2006dark}
\begin{barticle}
\bauthor{\bsnm{Lebel}, \binits{O.}},
\bauthor{\bsnm{Maris}, \binits{T.}},
\bauthor{\bsnm{Perron}, \binits{M.-{\`E}.}},
\bauthor{\bsnm{Demers}, \binits{E.}},
\bauthor{\bsnm{Wuest}, \binits{J.D.}}:
\batitle{The dark side of crystal engineering: creating glasses from small
  symmetric molecules that form multiple hydrogen bonds}.
\bjtitle{Journal of the American Chemical Society}
\bvolume{128}(\bissue{32}),
\bfpage{10372}--\blpage{10373}
(\byear{2006})
\end{barticle}
\endbibitem

\bibitem[\protect\citeauthoryear{Sheiko and
  Dobrynin}{2019}]{sheiko2019architectural}
\begin{barticle}
\bauthor{\bsnm{Sheiko}, \binits{S.S.}},
\bauthor{\bsnm{Dobrynin}, \binits{A.V.}}:
\batitle{Architectural code for rubber elasticity: from supersoft to superfirm
  materials}.
\bjtitle{Macromolecules}
\bvolume{52}(\bissue{20}),
\bfpage{7531}--\blpage{7546}
(\byear{2019})
\end{barticle}
\endbibitem

\bibitem[\protect\citeauthoryear{Asadi et~al.}{2023}]{asadi2023tuning}
\begin{barticle}
\bauthor{\bsnm{Asadi}, \binits{V.}},
\bauthor{\bsnm{Ruiz-Franco}, \binits{J.}},
\bauthor{\bsnm{Gucht}, \binits{J.}},
\bauthor{\bsnm{Kodger}, \binits{T.E.}}:
\batitle{Tuning moduli of hybrid bottlebrush elastomers by molecular
  architecture}.
\bjtitle{Materials \& Design}
\bvolume{234},
\bfpage{112326}
(\byear{2023})
\end{barticle}
\endbibitem

\bibitem[\protect\citeauthoryear{Lee and Ediger}{2010}]{lee2010mechanical}
\begin{barticle}
\bauthor{\bsnm{Lee}, \binits{H.-N.}},
\bauthor{\bsnm{Ediger}, \binits{M.}}:
\batitle{Mechanical rejuvenation in poly (methyl methacrylate) glasses?
  molecular mobility after deformation}.
\bjtitle{Macromolecules}
\bvolume{43}(\bissue{13}),
\bfpage{5863}--\blpage{5873}
(\byear{2010})
\end{barticle}
\endbibitem

\bibitem[\protect\citeauthoryear{Bennin et~al.}{2019}]{bennin2019enhanced}
\begin{barticle}
\bauthor{\bsnm{Bennin}, \binits{T.}},
\bauthor{\bsnm{Ricci}, \binits{J.}},
\bauthor{\bsnm{Ediger}, \binits{M.}}:
\batitle{Enhanced segmental dynamics of poly (lactic acid) glasses during
  constant strain rate deformation}.
\bjtitle{Macromolecules}
\bvolume{52}(\bissue{17}),
\bfpage{6428}--\blpage{6437}
(\byear{2019})
\end{barticle}
\endbibitem

\bibitem[\protect\citeauthoryear{Barr et~al.}{2023}]{barr2023autonomous}
\begin{barticle}
\bauthor{\bsnm{Barr}, \binits{C.M.}},
\bauthor{\bsnm{Duong}, \binits{T.}},
\bauthor{\bsnm{Bufford}, \binits{D.C.}},
\bauthor{\bsnm{Milne}, \binits{Z.}},
\bauthor{\bsnm{Molkeri}, \binits{A.}},
\bauthor{\bsnm{Heckman}, \binits{N.M.}},
\bauthor{\bsnm{Adams}, \binits{D.P.}},
\bauthor{\bsnm{Srivastava}, \binits{A.}},
\bauthor{\bsnm{Hattar}, \binits{K.}},
\bauthor{\bsnm{Demkowicz}, \binits{M.J.}}, \betal:
\batitle{Autonomous healing of fatigue cracks via cold welding}.
\bjtitle{Nature}
\bvolume{620}(\bissue{7974}),
\bfpage{552}--\blpage{556}
(\byear{2023})
\end{barticle}
\endbibitem

\bibitem[\protect\citeauthoryear{Wagle and Baker}{2015}]{wagle2015cold}
\begin{barticle}
\bauthor{\bsnm{Wagle}, \binits{D.V.}},
\bauthor{\bsnm{Baker}, \binits{G.A.}}:
\batitle{Cold welding: a phenomenon for spontaneous self-healing and shape
  genesis at the nanoscale}.
\bjtitle{Materials Horizons}
\bvolume{2}(\bissue{2}),
\bfpage{157}--\blpage{167}
(\byear{2015})
\end{barticle}
\endbibitem

\bibitem[\protect\citeauthoryear{Dai et~al.}{2016}]{dai2016side}
\begin{barticle}
\bauthor{\bsnm{Dai}, \binits{G.}},
\bauthor{\bsnm{Wang}, \binits{B.}},
\bauthor{\bsnm{Xu}, \binits{S.}},
\bauthor{\bsnm{Lu}, \binits{Y.}},
\bauthor{\bsnm{Shen}, \binits{Y.}}:
\batitle{Side-to-side cold welding for controllable nanogap formation from
  “dumbbell” ultrathin gold nanorods}.
\bjtitle{ACS Applied Materials \& Interfaces}
\bvolume{8}(\bissue{21}),
\bfpage{13506}--\blpage{13511}
(\byear{2016})
\end{barticle}
\endbibitem

\bibitem[\protect\citeauthoryear{Xu and Demkowicz}{2013}]{xu2013healing}
\begin{barticle}
\bauthor{\bsnm{Xu}, \binits{G.}},
\bauthor{\bsnm{Demkowicz}, \binits{M.}}:
\batitle{Healing of nanocracks by disclinations}.
\bjtitle{Physical review letters}
\bvolume{111}(\bissue{14}),
\bfpage{145501}
(\byear{2013})
\end{barticle}
\endbibitem

\bibitem[\protect\citeauthoryear{Giuntoli
  et~al.}{2020}]{giuntoli2020predictive}
\begin{barticle}
\bauthor{\bsnm{Giuntoli}, \binits{A.}},
\bauthor{\bsnm{Puosi}, \binits{F.}},
\bauthor{\bsnm{Leporini}, \binits{D.}},
\bauthor{\bsnm{Starr}, \binits{F.W.}},
\bauthor{\bsnm{Douglas}, \binits{J.F.}}:
\batitle{Predictive relation for the $\alpha$-relaxation time of a
  coarse-grained polymer melt under steady shear}.
\bjtitle{Science advances}
\bvolume{6}(\bissue{17}),
\bfpage{0777}
(\byear{2020})
\end{barticle}
\endbibitem

\bibitem[\protect\citeauthoryear{Ghosh and
  Schweizer}{2023}]{ghosh2023microscopic}
\begin{barticle}
\bauthor{\bsnm{Ghosh}, \binits{A.}},
\bauthor{\bsnm{Schweizer}, \binits{K.S.}}:
\batitle{Microscopic activated dynamics theory of the shear rheology and stress
  overshoot in ultradense glass-forming fluids and colloidal suspensions}.
\bjtitle{Journal of Rheology}
\bvolume{67}(\bissue{2}),
\bfpage{559}--\blpage{578}
(\byear{2023})
\end{barticle}
\endbibitem

\bibitem[\protect\citeauthoryear{Pham et~al.}{2008}]{pham2008yielding}
\begin{barticle}
\bauthor{\bsnm{Pham}, \binits{K.}},
\bauthor{\bsnm{Petekidis}, \binits{G.}},
\bauthor{\bsnm{Vlassopoulos}, \binits{D.}},
\bauthor{\bsnm{Egelhaaf}, \binits{S.}},
\bauthor{\bsnm{Poon}, \binits{W.}},
\bauthor{\bsnm{Pusey}, \binits{P.}}:
\batitle{Yielding behavior of repulsion-and attraction-dominated colloidal
  glasses}.
\bjtitle{Journal of Rheology}
\bvolume{52}(\bissue{2}),
\bfpage{649}--\blpage{676}
(\byear{2008})
\end{barticle}
\endbibitem

\bibitem[\protect\citeauthoryear{Koumakis and
  Petekidis}{2011}]{koumakis2011two}
\begin{barticle}
\bauthor{\bsnm{Koumakis}, \binits{N.}},
\bauthor{\bsnm{Petekidis}, \binits{G.}}:
\batitle{Two step yielding in attractive colloids: transition from gels to
  attractive glasses}.
\bjtitle{Soft Matter}
\bvolume{7}(\bissue{6}),
\bfpage{2456}--\blpage{2470}
(\byear{2011})
\end{barticle}
\endbibitem

\bibitem[\protect\citeauthoryear{Moghimi and
  Petekidis}{2020}]{moghimi2020mechanisms}
\begin{barticle}
\bauthor{\bsnm{Moghimi}, \binits{E.}},
\bauthor{\bsnm{Petekidis}, \binits{G.}}:
\batitle{Mechanisms of two-step yielding in attractive colloidal glasses}.
\bjtitle{Journal of Rheology}
\bvolume{64}(\bissue{5}),
\bfpage{1209}--\blpage{1225}
(\byear{2020})
\end{barticle}
\endbibitem

\bibitem[\protect\citeauthoryear{Regev et~al.}{2013}]{regev2013onset}
\begin{barticle}
\bauthor{\bsnm{Regev}, \binits{I.}},
\bauthor{\bsnm{Lookman}, \binits{T.}},
\bauthor{\bsnm{Reichhardt}, \binits{C.}}:
\batitle{Onset of irreversibility and chaos in amorphous solids under periodic
  shear}.
\bjtitle{Physical Review E}
\bvolume{88}(\bissue{6}),
\bfpage{062401}
(\byear{2013})
\end{barticle}
\endbibitem

\bibitem[\protect\citeauthoryear{Leishangthem
  et~al.}{2017}]{leishangthem2017yielding}
\begin{barticle}
\bauthor{\bsnm{Leishangthem}, \binits{P.}},
\bauthor{\bsnm{Parmar}, \binits{A.D.}},
\bauthor{\bsnm{Sastry}, \binits{S.}}:
\batitle{The yielding transition in amorphous solids under oscillatory shear
  deformation}.
\bjtitle{Nature communications}
\bvolume{8}(\bissue{1}),
\bfpage{14653}
(\byear{2017})
\end{barticle}
\endbibitem

\bibitem[\protect\citeauthoryear{Schinasi-Lemberg and
  Regev}{2020}]{schinasi2020annealing}
\begin{barticle}
\bauthor{\bsnm{Schinasi-Lemberg}, \binits{E.}},
\bauthor{\bsnm{Regev}, \binits{I.}}:
\batitle{Annealing and rejuvenation in a two-dimensional model amorphous solid
  under oscillatory shear}.
\bjtitle{Physical Review E}
\bvolume{101}(\bissue{1}),
\bfpage{012603}
(\byear{2020})
\end{barticle}
\endbibitem

\bibitem[\protect\citeauthoryear{Yeh et~al.}{2020}]{yeh2020glass}
\begin{barticle}
\bauthor{\bsnm{Yeh}, \binits{W.-T.}},
\bauthor{\bsnm{Ozawa}, \binits{M.}},
\bauthor{\bsnm{Miyazaki}, \binits{K.}},
\bauthor{\bsnm{Kawasaki}, \binits{T.}},
\bauthor{\bsnm{Berthier}, \binits{L.}}:
\batitle{Glass stability changes the nature of yielding under oscillatory
  shear}.
\bjtitle{Physical review letters}
\bvolume{124}(\bissue{22}),
\bfpage{225502}
(\byear{2020})
\end{barticle}
\endbibitem

\bibitem[\protect\citeauthoryear{Bhaumik et~al.}{2022}]{bhaumik2022yielding}
\begin{botherref}
\oauthor{\bsnm{Bhaumik}, \binits{H.}},
\oauthor{\bsnm{Foffi}, \binits{G.}},
\oauthor{\bsnm{Sastry}, \binits{S.}}:
Yielding transition of a two dimensional glass former under athermal cyclic
  shear deformation.
The Journal of Chemical Physics
\textbf{156}(6)
(2022)
\end{botherref}
\endbibitem

\bibitem[\protect\citeauthoryear{Parmar et~al.}{2019}]{parmar2019strain}
\begin{barticle}
\bauthor{\bsnm{Parmar}, \binits{A.D.}},
\bauthor{\bsnm{Kumar}, \binits{S.}},
\bauthor{\bsnm{Sastry}, \binits{S.}}:
\batitle{Strain localization above the yielding point in cyclically deformed
  glasses}.
\bjtitle{Physical Review X}
\bvolume{9}(\bissue{2}),
\bfpage{021018}
(\byear{2019})
\end{barticle}
\endbibitem

\bibitem[\protect\citeauthoryear{Chen and Schweizer}{2011}]{chen2011theory}
\begin{barticle}
\bauthor{\bsnm{Chen}, \binits{K.}},
\bauthor{\bsnm{Schweizer}, \binits{K.S.}}:
\batitle{Theory of yielding, strain softening, and steady plastic flow in
  polymer glasses under constant strain rate deformation}.
\bjtitle{Macromolecules}
\bvolume{44}(\bissue{10}),
\bfpage{3988}--\blpage{4000}
(\byear{2011})
\end{barticle}
\endbibitem

\bibitem[\protect\citeauthoryear{Bhaumik et~al.}{2021}]{bhaumik2021role}
\begin{barticle}
\bauthor{\bsnm{Bhaumik}, \binits{H.}},
\bauthor{\bsnm{Foffi}, \binits{G.}},
\bauthor{\bsnm{Sastry}, \binits{S.}}:
\batitle{The role of annealing in determining the yielding behavior of glasses
  under cyclic shear deformation}.
\bjtitle{Proceedings of the National Academy of Sciences}
\bvolume{118}(\bissue{16}),
\bfpage{2100227118}
(\byear{2021})
\end{barticle}
\endbibitem

\bibitem[\protect\citeauthoryear{Bhaumik et~al.}{2022}]{bhaumik2022avalanches}
\begin{barticle}
\bauthor{\bsnm{Bhaumik}, \binits{H.}},
\bauthor{\bsnm{Foffi}, \binits{G.}},
\bauthor{\bsnm{Sastry}, \binits{S.}}:
\batitle{Avalanches, clusters, and structural change in cyclically sheared
  silica glass}.
\bjtitle{Physical Review Letters}
\bvolume{128}(\bissue{9}),
\bfpage{098001}
(\byear{2022})
\end{barticle}
\endbibitem

\bibitem[\protect\citeauthoryear{Ozawa et~al.}{2018}]{ozawa2018random}
\begin{barticle}
\bauthor{\bsnm{Ozawa}, \binits{M.}},
\bauthor{\bsnm{Berthier}, \binits{L.}},
\bauthor{\bsnm{Biroli}, \binits{G.}},
\bauthor{\bsnm{Rosso}, \binits{A.}},
\bauthor{\bsnm{Tarjus}, \binits{G.}}:
\batitle{Random critical point separates brittle and ductile yielding
  transitions in amorphous materials}.
\bjtitle{Proceedings of the National Academy of Sciences}
\bvolume{115}(\bissue{26}),
\bfpage{6656}--\blpage{6661}
(\byear{2018})
\end{barticle}
\endbibitem

\bibitem[\protect\citeauthoryear{Shrivastav
  et~al.}{2016}]{shrivastav2016yielding}
\begin{barticle}
\bauthor{\bsnm{Shrivastav}, \binits{G.P.}},
\bauthor{\bsnm{Chaudhuri}, \binits{P.}},
\bauthor{\bsnm{Horbach}, \binits{J.}}:
\batitle{Yielding of glass under shear: A directed percolation transition
  precedes shear-band formation}.
\bjtitle{Physical Review E}
\bvolume{94}(\bissue{4}),
\bfpage{042605}
(\byear{2016})
\end{barticle}
\endbibitem

\bibitem[\protect\citeauthoryear{Larini et~al.}{2008}]{larini2008universal}
\begin{barticle}
\bauthor{\bsnm{Larini}, \binits{L.}},
\bauthor{\bsnm{Ottochian}, \binits{A.}},
\bauthor{\bsnm{De~Michele}, \binits{C.}},
\bauthor{\bsnm{Leporini}, \binits{D.}}:
\batitle{Universal scaling between structural relaxation and vibrational
  dynamics in glass-forming liquids and polymers}.
\bjtitle{Nature Physics}
\bvolume{4}(\bissue{1}),
\bfpage{42}--\blpage{45}
(\byear{2008})
\end{barticle}
\endbibitem

\bibitem[\protect\citeauthoryear{Zhu et~al.}{2022}]{zhu2022effect}
\begin{botherref}
\oauthor{\bsnm{Zhu}, \binits{Y.}},
\oauthor{\bsnm{Giuntoli}, \binits{A.}},
\oauthor{\bsnm{Zhang}, \binits{W.}},
\oauthor{\bsnm{Lin}, \binits{Z.}},
\oauthor{\bsnm{Keten}, \binits{S.}},
\oauthor{\bsnm{Starr}, \binits{F.W.}},
\oauthor{\bsnm{Douglas}, \binits{J.F.}}:
The effect of nanoparticle softness on the interfacial dynamics of a model
  polymer nanocomposite.
The Journal of Chemical Physics
\textbf{157}(9)
(2022)
\end{botherref}
\endbibitem

\bibitem[\protect\citeauthoryear{Fiocco et~al.}{2014}]{fiocco2014encoding}
\begin{barticle}
\bauthor{\bsnm{Fiocco}, \binits{D.}},
\bauthor{\bsnm{Foffi}, \binits{G.}},
\bauthor{\bsnm{Sastry}, \binits{S.}}:
\batitle{Encoding of memory in sheared amorphous solids}.
\bjtitle{Physical review letters}
\bvolume{112}(\bissue{2}),
\bfpage{025702}
(\byear{2014})
\end{barticle}
\endbibitem

\bibitem[\protect\citeauthoryear{Adhikari and
  Sastry}{2018}]{adhikari2018memory}
\begin{barticle}
\bauthor{\bsnm{Adhikari}, \binits{M.}},
\bauthor{\bsnm{Sastry}, \binits{S.}}:
\batitle{Memory formation in cyclically deformed amorphous solids and sphere
  assemblies}.
\bjtitle{The European Physical Journal E}
\bvolume{41},
\bfpage{1}--\blpage{17}
(\byear{2018})
\end{barticle}
\endbibitem

\bibitem[\protect\citeauthoryear{Thompson et~al.}{2022}]{LAMMPS}
\begin{barticle}
\bauthor{\bsnm{Thompson}, \binits{A.P.}},
\bauthor{\bsnm{Aktulga}, \binits{H.M.}},
\bauthor{\bsnm{Berger}, \binits{R.}},
\bauthor{\bsnm{Bolintineanu}, \binits{D.S.}},
\bauthor{\bsnm{Brown}, \binits{W.M.}},
\bauthor{\bsnm{Crozier}, \binits{P.S.}},
\bauthor{\bsnm{Veld}, \binits{P.J.}},
\bauthor{\bsnm{Kohlmeyer}, \binits{A.}},
\bauthor{\bsnm{Moore}, \binits{S.G.}},
\bauthor{\bsnm{Nguyen}, \binits{T.D.}},
\bauthor{\bsnm{Shan}, \binits{R.}},
\bauthor{\bsnm{Stevens}, \binits{M.J.}},
\bauthor{\bsnm{Tranchida}, \binits{J.}},
\bauthor{\bsnm{Trott}, \binits{C.}},
\bauthor{\bsnm{Plimpton}, \binits{S.J.}}:
\batitle{{LAMMPS} - a flexible simulation tool for particle-based materials
  modeling at the atomic, meso, and continuum scales}.
\bjtitle{Comp. Phys. Comm.}
\bvolume{271},
\bfpage{108171}
(\byear{2022})
\doiurl{10.1016/j.cpc.2021.108171}
\end{barticle}
\endbibitem

\bibitem[\protect\citeauthoryear{Stukowski}{2009}]{stukowski2009visualization}
\begin{barticle}
\bauthor{\bsnm{Stukowski}, \binits{A.}}:
\batitle{Visualization and analysis of atomistic simulation data with
  ovito--the open visualization tool}.
\bjtitle{Modelling and simulation in materials science and engineering}
\bvolume{18}(\bissue{1}),
\bfpage{015012}
(\byear{2009})
\end{barticle}
\endbibitem

\end{thebibliography}

\clearpage


\section*{\centering Supplemental Material for \\ ``Inducing mechanical self-healing in glassy polymer melts''}
\setcounter{equation}{0}
\setcounter{figure}{0}
\setcounter{table}{0}
\setcounter{section}{0}

\renewcommand{\theequation}{S\arabic{equation}}
\renewcommand{\thefigure}{S\arabic{figure}}

\tableofcontents

\newpage

\section{\centering Static and dynamics properties in equilibrium}
\label{sec:s1}
\begin{figure}[th]
\includegraphics[width=\linewidth]{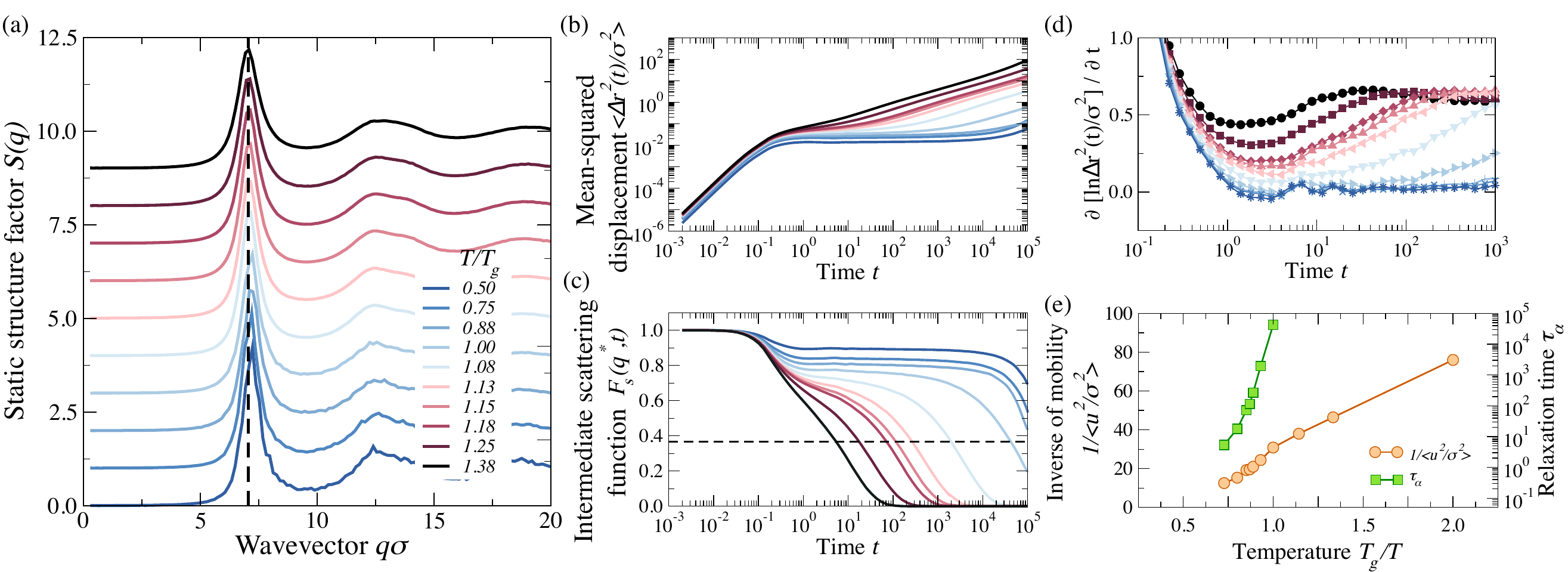}
\caption{\label{Sta_Dyn} Static and dynamic properties as a function of temperature $T$. (a) Static structure factor $S\left(q\right)$ shifted in the y-axis by an arbitrary factor to improve the visualization. The vertical dashed line highlights the position of the peak $q^{*}$. (b) Mean-squared displacement $\left\langle \Delta r^{2}\left(t\right) \right\rangle$. (c) Incoherent intermediate scattering function $F_{s}\left(q^{*},t\right)$. The horizontal dashed line corresponds to $F_{s}\left(q^{*},\tau_{\alpha}\right)=e^{-1}$, $\tau_{\alpha}$ being the structural relaxation time. (d) Logarithmic derivative of the mean-squared displacement $\partial\left[ ln r^{2}\left(t\right) \right] /\partial ln\,t $ as a function of $t$. (e) Relaxation time $\tau_{\alpha}$ (on the right) and mobility $\left\langle u^{2} \right\rangle$ (on the left) as a function of $T_{g}/T$.}
\end{figure}

Fig.~\ref{Sta_Dyn} collects the static and dynamic properties of the glassy polymer melts as a function of temperature $T$. In particular, Fig.~\ref{Sta_Dyn}(a) shows the static structure factor $S\left(q\right)$, where the position of the main peak is at $q^{*} \sim 7.14$, independently of $T$. Fig.~\ref{Sta_Dyn}(b) and (c) display the mean-squared displacement (MSD) $\left\langle \Delta r^{2}\left(t\right) \right\rangle$ and self-intermediate scattering function $F_{s}\left(q^{*},t\right)$, respectively. While at short times we observe ballistic dynamics, the MSD develops a plateau with decreasing $T$, revealing the onset of the caging dynamics. This plateau extends for longer times with decreasing $T$, corresponding to an increased caging of the particles, until $T \sim T_g$, at which point the relaxation time becomes comparable to the time window of observation. Likewise, the onset of the glassy state is observed through the $F_{s}\left(q^{*},t\right)$, which exhibits a double decay with decreasing $T$. The first decay, observed at short times, is a consequence of the interactions of the particles with their neighbors that form cages ($\beta$-relaxation). The second decay, taking place at longer times, indicates the $\alpha-$relaxation of the system at a time $\tau_{\alpha}$ which we define as $F_{s}\left( q^{*},\tau_{\alpha} \right)=1/e$. The particles' degree of mobility can be also characterized by the Debye-Waller factor $\left\langle u^{2} \right\rangle$, initially associated with the mean-square amplitude of atoms in the solid state around their equilibrium positions, and later, extrapolated to glassy states to quantify the size of cages. We compute the quantity $\partial ln\left[ r^{2}\left(t\right) \right] /\partial ln\,t $, shown in Fig.~\ref{Sta_Dyn}(d). The minimum corresponding to the inflection point in the log-log plot of MSD estimates a characteristic time of $\beta$-relaxations $t_\beta$~\cite{larini2008universal,zhu2022effect}. Knowing this time, the Debye-Waller factor is defined as $\left\langle u^{2} \right\rangle = \left\langle \Delta r^{2}\left( t_{\beta} \right) \right\rangle$. Finally, in Fig.~\ref{Sta_Dyn}(e), we report the mobility and relaxation time as a function of $T_{g}/T$. The increase in $\tau_{\alpha}$ with decreasing $T$ is attributed to the particles becoming more localized within microscopic cages, as indicated by $\left\langle u^{2} \right\rangle$.

\newpage

\section{\centering Rheological behavior under steady shear flow}
\label{sec:s2}

\begin{figure}[th]
\includegraphics[width=\linewidth]{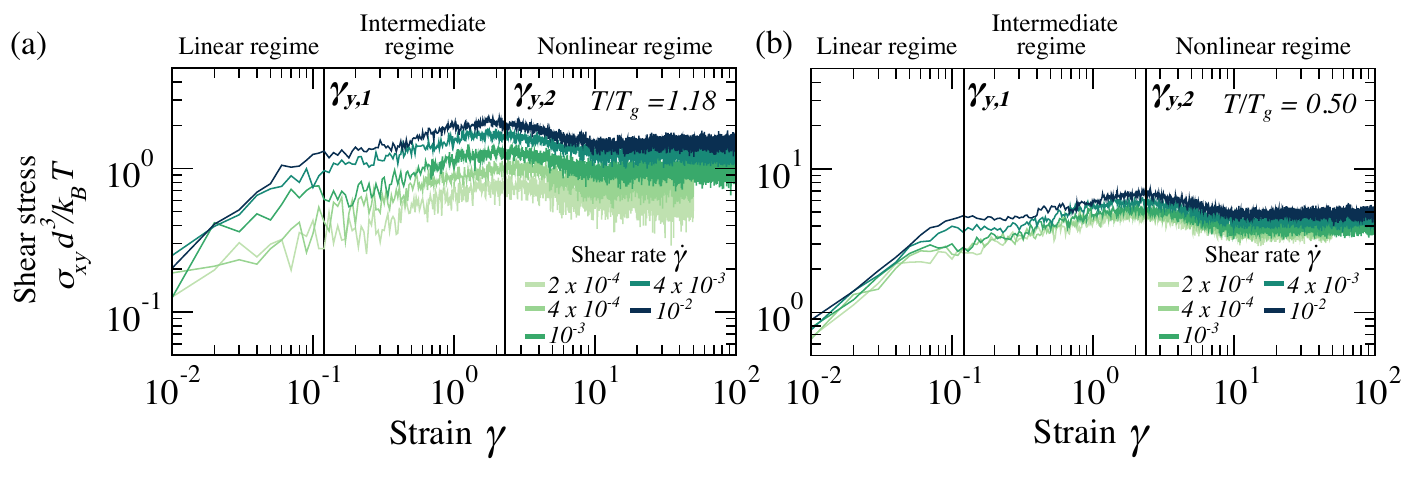}
\caption{\label{Steady} Shear stress tensor $\sigma_{xy} d^{3}/k_{B}T$ versus strain $\gamma$ as a function of the shear rate $\dot{\gamma}$ at (a) $T/T_{g}=1.18$ and (b) $T/T_{g}=0.50$ as a function of $T$ at shear rate $\dot{\gamma} = 0.01$. Vertical black lines highlight the bond-breaking point $\gamma_{y,1}$ and the yielding point $\gamma_{y,2}$ corresponding to the transition from solid-like to fluid-like behavior.}
\end{figure}

We apply a steady shear deformation, following the protocol explained in \textit{Methods} section, to study the rheological behavior of our glassy polymer melts. As explained in Fig. 1(a) from the main, polymer melts exhibit a transition from a solid-like behavior at small strains to a liquid-like behavior at large strains. This transition is highlighted by the stress overshot at the yielding point $\gamma_{y,2}$. This point indicates the maximum stress that the system can accumulate before the solid-to-liquid transition. The height of the yielding point increases with decreasing $T$, corresponding to the presence of a stronger inherent structure able to support more stress. Likewise, the position of the yielding point is not temperature-dependent. Thus, the viscoelastic transition is determined by the breakdown of geometric frustration emerging from the confinement of particles within cages. However, we also see in Fig.~\ref{Steady}(a) and (b) that with the decrease of $T$, a first yielding point emerges at small strains. This yielding point $\gamma_{y,1}$ was previously reported in attractive colloidal glasses~\cite{pham2008yielding,koumakis2011two,moghimi2020mechanisms}, and related to the breaking of physical bonds. Since we are considering attractive interactions between non-bonded monomers, we assert that $\gamma_{y,1}$ corresponds to the same bond-breaking mechanisms. Furthermore, we observe that the $\sigma_{xy}\left(\gamma_{b,1}\right)$ depends on $\dot{\gamma}$.

\newpage

\section{\centering Oscillatory deformations for glassy polymer melts}
\label{sec:s3}

\subsection{\centering Self-intermediate scattering function under oscillatory shear}

\begin{figure}[!h]
\includegraphics[width=0.9\linewidth]{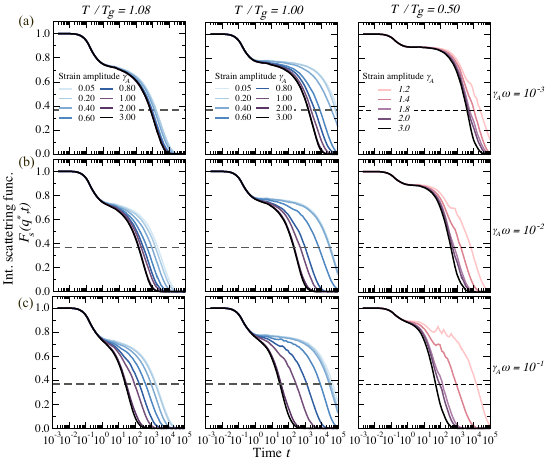}
\caption{\label{S3}  Incoherent intermediate scattering function $F_{s}\left(q^{*},t\right)$ under oscillatory shear flow as a function of the strain amplitude $\gamma_{A}$ and for (a) $\gamma_{A}\omega=10^{-3}$, (b) $\gamma_{A}\omega=10^{-2}$, and (c) $\gamma_{A}\omega=10^{-1}$ for polymer melts at $T/T_{g}=1.08$, $T/T_{g} = 1.00$, and $T/T_{g}=0.50$. The horizontal dashed line corresponds to $F_{s}\left(q^{*},\tau_{\alpha}\right)=e^{-1}$, where $\tau_{\alpha}$ is the structural relaxation time.}
\end{figure}

Fig.~\ref{S3} shows $F_{s}\left(q^{*},t\right)$ under oscillatory deformations at different $T$, strain amplitudes, and deformation rate, computed at length scale distance $q^{*}=2\pi/l=7.14$. Since oscillatory deformations are applied in the $xy-$plane, with the $x-$axis representing the velocity direction, the intermediate scattering function is computed by excluding the $x$ component of the wave vector $\overline{q^{*}}$. In each panel, $F_{s}\left(q^{*},t\right)$ is represented as a function of the strain amplitude $\gamma_{A}$. Results reveal that by increasing the strain amplitude $\gamma_{A}$ and frequency $\omega$, the structural relaxation time $\tau_{\alpha}$ shifts to shorter times.

\newpage

\subsection{\centering Relaxation time of polymer melts at $T > T_{g}$}

\begin{figure}[!h]
\includegraphics[width=\linewidth]{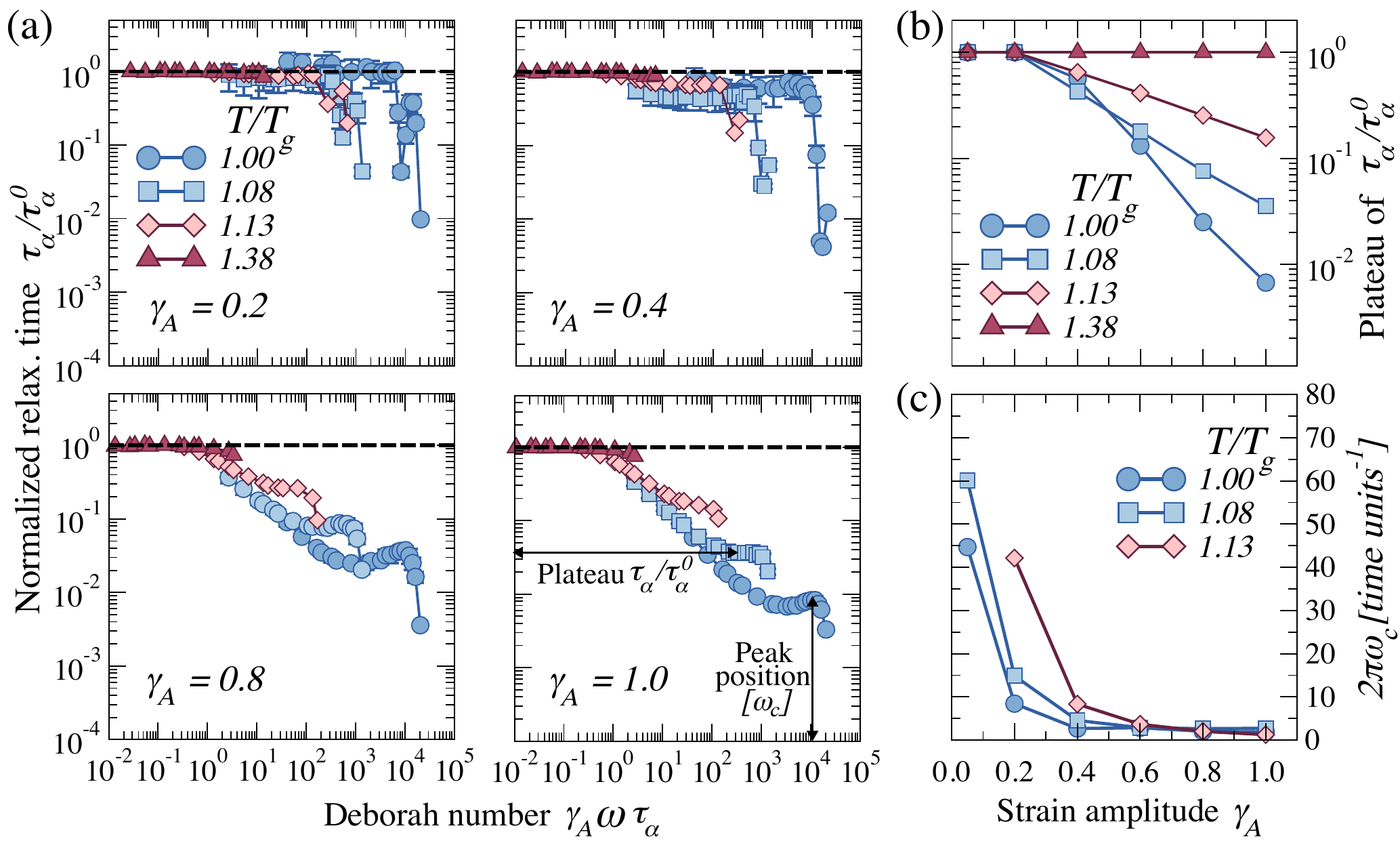}
\caption{\label{S4}  (a) Normalized relaxation time $\tau_{\alpha}/\tau_{\alpha}^{0}$ as a function of the Deborah number $\gamma_{A}\omega\tau_{\alpha}^{0}$, for systems at different $T$ subjected to oscillatory deformations. Here, $\tau_{\alpha}^{0}$ refers to the relaxation time at rest. The linear regime corresponds to deformations with $\gamma_{A}<\gamma_{y,1}$, the intermediate regime encompasses the range $\gamma_{y,1}<\gamma_{A}<\gamma_{y,2}$, and the nonlinear regime is found at $\gamma_{A}>\gamma_{y,2}$. The horizontal dashed line emphasizes the changes in the system's dynamics originated from the deformation. Additional arrows are represented in the last panel to highlight the plateau height and peak position $\omega_{c}$, respectively. (b) Plateau value of $\tau_{\alpha}/\tau_{\alpha}^{0}$ and (c) position of the peak as a function of $\gamma_{A}$ and $T/T_{g}$.}
\end{figure}

In Fig.~\ref{S4}(a) we show the normalized relaxation time $\tau_{\alpha}/\tau_{\alpha}^{0}$ as a function of Deborah number $\gamma_{A}\omega\tau_{\alpha}^{0}$, in the intermediate regime $\gamma_{y,1} < \gamma_{A} < \gamma_{y,2}$. As discussed in the main, when Deborah numbers are larger than $1$ the dynamics accelerates until reaching a plateau. The plateau length increases as we approach $T_{g}$. On the other hand, $\tau_{\alpha}$ falls to a plateau of lower height as $\gamma_{A}$ approaches $\gamma_{y,2}$, see Fig.~\ref{S4}(b). Furthermore, as $\gamma_{A}\omega$ increases with increasing $\omega$, the system approaches a critical value $\omega_{c}$ after which $\tau_{\alpha}$ sharply decreases. In Fig.~\ref{S4}(c) we report that $\omega_{c}$ decays exponentially with increasing $\gamma_{A}$. Consistently, the peak disappears in the nonlinear regime. The presence of this maximum following a sharp decrease in $\tau_{\alpha}$ suggests a non-trivial annealing behavior of the glasses near the yielding point~\cite{parmar2019strain}. However, as discussed below in Fig.~\ref{S7}, evidence of this is not observed in the distribution of per-particle mobility.

\newpage

\subsection{\centering Energy per particle as a function of oscillatory deformations}

\begin{figure}[!h]
\includegraphics[width=\linewidth]{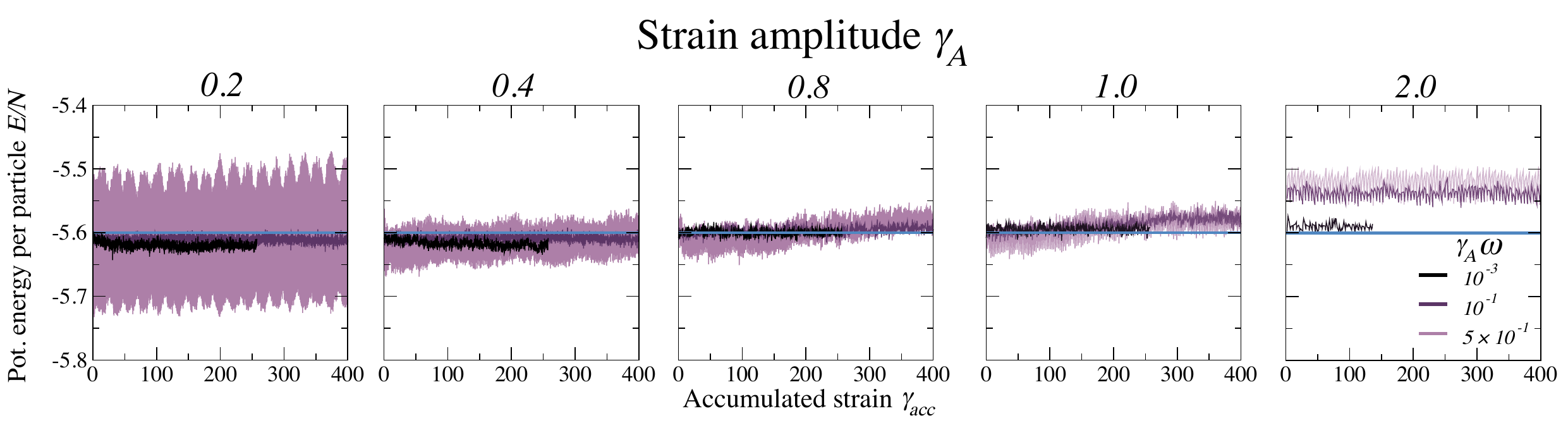}
\caption{\label{Energy}  Energy per particle $E/N$ as a function of accumulated strain $\gamma_{acc}$ and $\gamma_{A}\omega$ for a system at $T/T_{g}=1.00$ under oscillatory deformations in the intermediate regime. The horizontal blue line corresponds to the energy value immediately before applying any deformation.}
\end{figure}

We represent in Fig.~\ref{Energy} the evolution of the energy per particle $E/N$, in the range $\gamma_{y,1}<\gamma_{A}<\gamma_{y,2}$, as a function of the accumulated strain $\gamma_{acc}$. While $E/N$ changes smoothly for all $\gamma_{A}$ values, without exhibiting sharp fluctuations associated with the formation of a shear banding, we observe significantly larger variations in energy for $\gamma_{A}=0.2$ and $\gamma_{A}\omega=0.5$. Given the proximity of $\gamma_{A}$ to the linear regime, the applied deformation could result in the system oscillating as a whole.

\subsection{\centering Particle displacement}

\begin{figure}[!h]
\includegraphics[width=\linewidth]{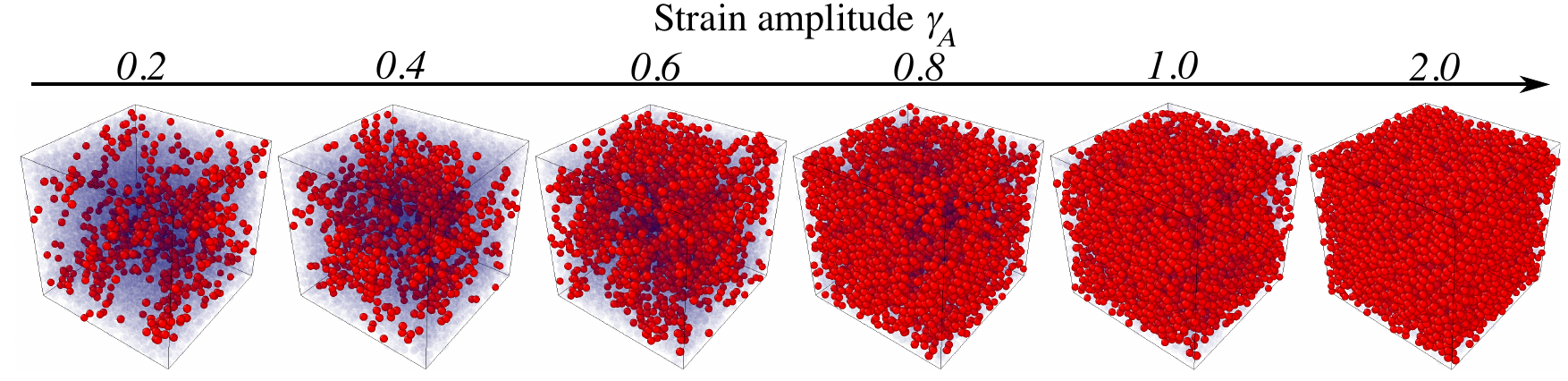}
\caption{\label{S6}  Snapshots for a glassy polymer melt at $T/T_{g}=1.00$ with accumulated strain $\gamma_{acc}=100$, as a function of strain amplitude $\gamma_{A}$. Here, red beads highlight particles that are displaced equal to or more than a distance equivalent to the particle size $d$ due to the accumulated deformation.}
\end{figure}

In Fig.~\ref{S6}, red beads highlight particles that are displaced by a distance equal to or greater than the particle size $d$ after having withstood $\gamma_{acc}=100\%$. This analysis was carried out as a function of $\gamma_{A}$. This finding, along with the smooth variation of $E/N$, indicates that the acceleration of the local dynamics in our study is not attributed to the presence of shear banding.

\newpage

\subsection{\centering Molecular mobility for polymer melts at $T < T_{g}$}

\begin{figure}[!h]
\includegraphics[width=\linewidth]{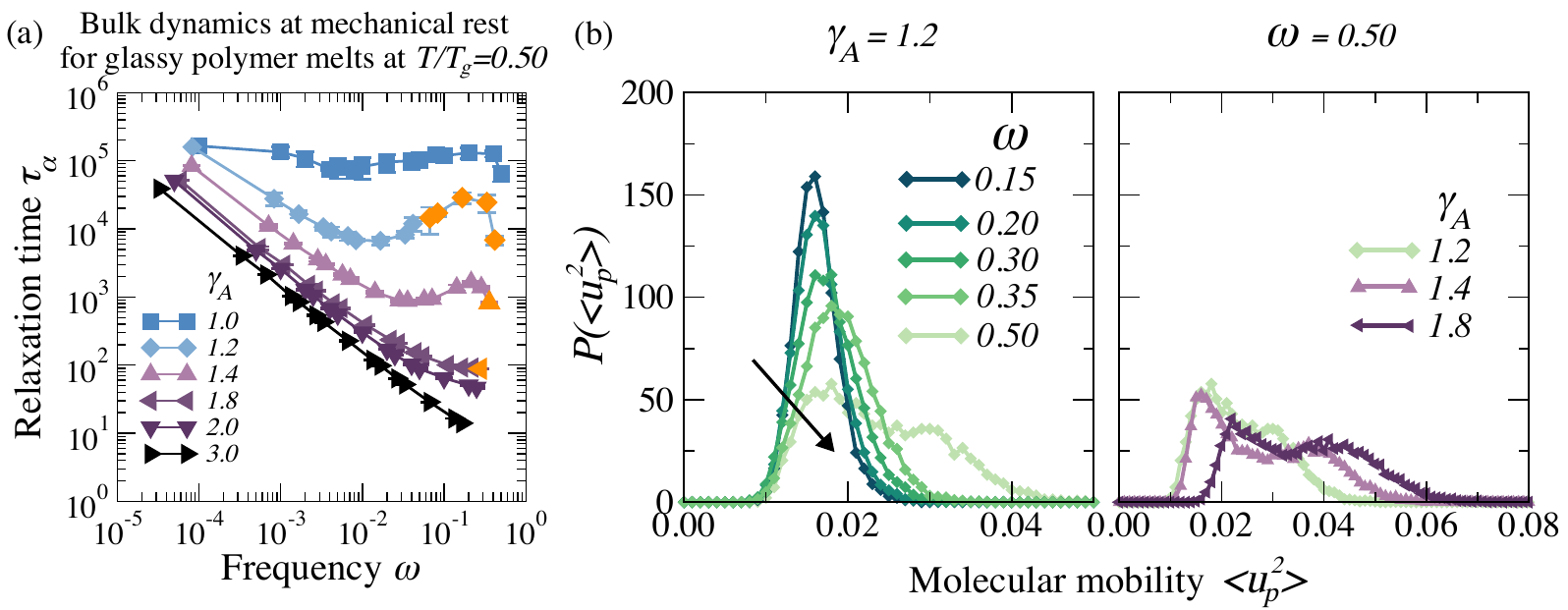}
\caption{\label{S7}  (a) Structural relaxation time $\tau_{\alpha}$ as a function of amplitude $\gamma_{A}$ and frequency $\omega$ deformation for a glassy polymer melt.
(b) Distribution of per-particle molecular mobility $P\left(\left\langle u_{p}^{2} \right\rangle \right)$, computed for the state points corresponding to the $\tau_{\alpha}$ values highlighted in panel (a). The arrow refers to the increase in frequency.}
\end{figure}

As discussed in the main text, Fig.~\ref{S7}(a) shows that oscillatory deformations in the range $\gamma_{y,1}<\gamma_{A}<\gamma_{y,2}$ can locally accelerate the system dynamics by breaking microscopic cages responsible for the material vitrification. To support this statement, we represent in Fig.~\ref{S7}(b) the distribution of per-particle Debye-Waller factor $<u_{p}^{2}>$. Fixing $\gamma_{A}$ and tuning $\omega$, $P\left(\left\langle u_{p}^{2} \right\rangle \right)$ shifts slightly to higher mobility. However, the increase in $\tau_{\alpha}$ at the critical $\omega_{c}$ does not correspond to a larger number of slow particles. In contrast, it results in a distribution with a long tail, indicating the formation of a wide spectrum of molecular mobility. The manifestation of this peak could be related to some type of anomalous diffusion in the system, which cannot be captured by molecular mobility. Further studies in this direction are needed. Finally, we see from Fig.~\ref{S7}(b) that the position and width of $P\left(\left\langle u_{p}^{2} \right\rangle \right)$ increases with $\gamma_{A}$, in agreement with the trend exhibited by $\tau_{\alpha}$.

\section{\centering Mechanical self-healing}

\subsection{\centering Crack evolution in equilibrium and under oscillatory deformations}
\label{sec:s5}

\begin{figure}[!h]
\includegraphics[width=\linewidth]{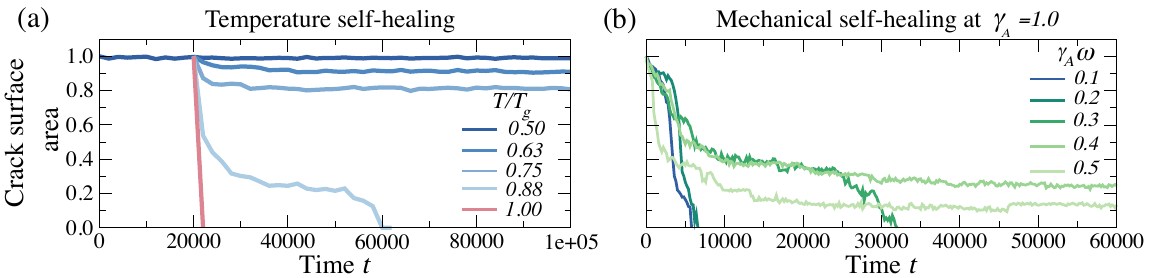}
\caption{\label{crack}  Crack surface area as a function of time $t$ for a cylindrical crack with initial diameter $D/d=1.5$. (a) Self-healing induced by increasing temperature $T$. A(b) Mechanical self-healing induced by applying oscillatory deformations with amplitude $\gamma_{A}=1.0$.}
\end{figure}

Fig.~\ref{crack} displays a specific sample with an initial diameter $D/d=1.5$, subjected to a general increase in $T$ while in mechanical equilibrium, see Fig.~\ref{crack}(a), or to oscillatory deformations of amplitude $\gamma_{A}=1.0$ at fixed temperature, see Fig.~\ref{crack}(b). Crack closure is reached when the crack surface is zero. Supplementary Video 1 shows the crack closure showing the displacement of the beads, while Supplementary Video 2 displays the evolution of the crack during the oscillatory deformation for $\gamma_{A}=1.0$ and $\gamma_{A}\omega=0.3$.

From Fig.~\ref{crack}(b), we observe that increasing $\omega$ leads to longer times for mechanical self-healing. This observation is summarized in phase diagrams discussed in Fig. 2(d) in the main text, where we observe mechanical self-healing begins to fail at high values of $\omega$. This is attributed to the fact that the oscillation rate of the deformation imposed by $\omega$ becomes faster than the characteristic particle diffusion rate. As a consequence, particles will follow the same trajectory in each deformation cycle, leading to a delay in the closure of the crack.

\subsection{\centering Local mobility}
\label{sec:s6}

\begin{figure}[!h]
\includegraphics[width=\linewidth]{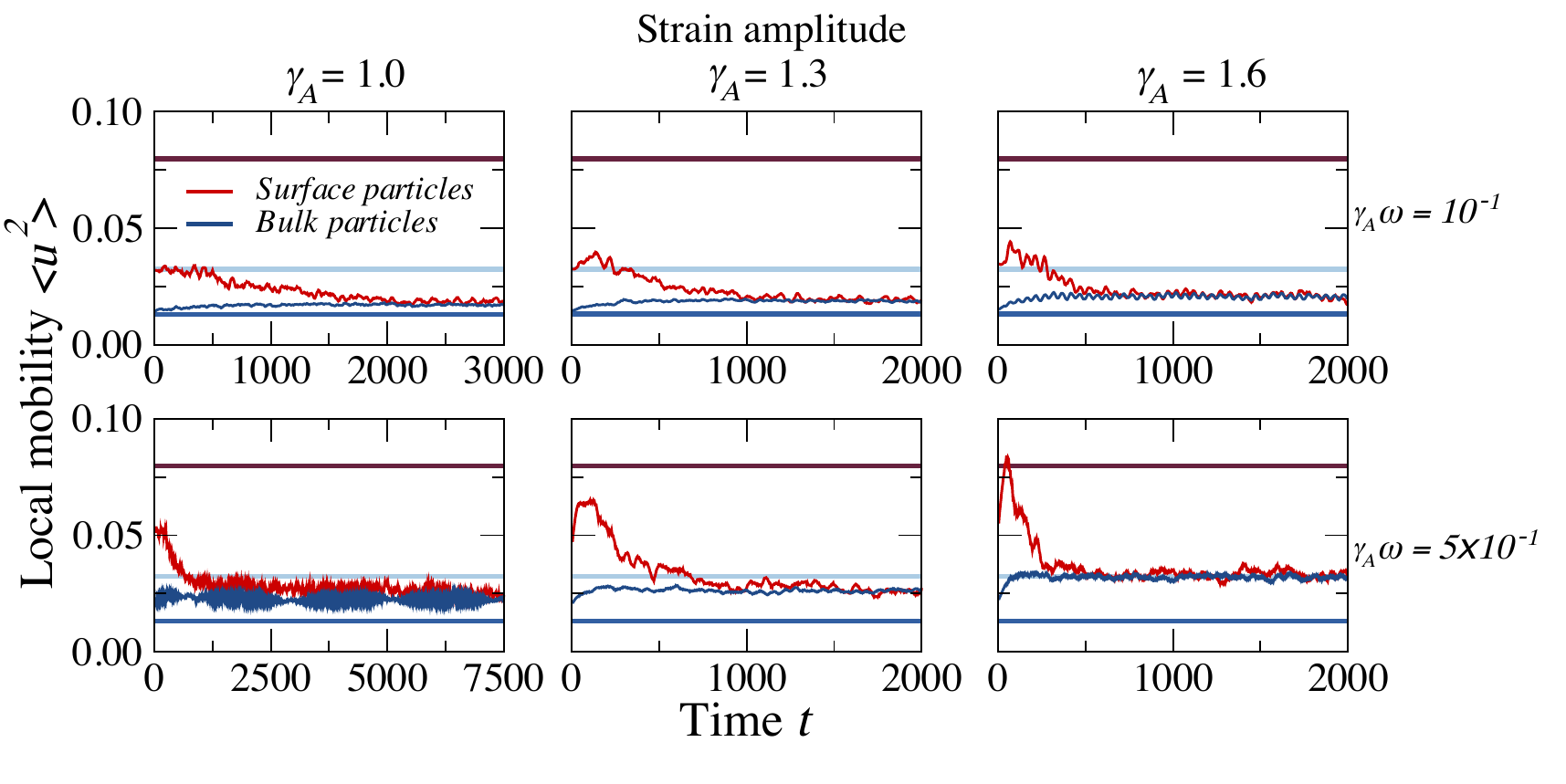}
\caption{\label{locmob}  Temporal evolution of the local mobility $\left\langle u^{2} \right\rangle$ for particles on the surface of a crack with diameter $D/d=1.0$ and in the bulk of a glassy polymer melt as a function of the amplitude oscillations $\gamma_{A}$ and the shear peak $\gamma_{A}\omega$. Curves were filtered with the Savitzky-Golay filter on windows of 21 points and equations of 2 order. Horizontal dark blue, light blue, and dark red lines correspond to the local mobility for $T/T_{g}=0.50$, $1.00$, and $1.38$, respectively.}
\end{figure}

Figure~\ref{locmob} illustrates the evolution of particle mobility $\left\langle u^{2} \right\rangle$ for particles on the crack surface with a diameter of $D/d=1.0$, compared to those within the bulk of a glassy polymer melt. These results confirm that the application of oscillatory deformations accurately accelerates the local mobility of the crack surface, only slightly modifying bulk mobility. Indeed, the crack surface can fully melt, exhibiting faster dynamics than bulk particles of the system at rest at $T>>T_{g}$. These findings, along with the discussion of Fig. 3 in the main text, prove the significant utility of oscillatory deformations in inducing self-healing.

\end{document}